\documentclass[12pt]{article}  
\usepackage{epsfig,epsf}  
 1 
\def\thefootnote{\fnsymbol{footnote}} 
\def\as{\alpha_{\rm s}}

\def\dd{\partial}

\def\percent{{\%\ }} 
 
\newcommand{\half}{\mbox{\small $\frac{1}{2}$}} 
 
\def\lsim{\;\raisebox{-.4ex}{\rlap{$\sim$}} \raisebox{.4ex}{$<$}\;}

\def\gevs{{\,\mbox{GeV}^2}} 
\def\gevm{{\,\mbox{GeV}^{-1}}} 
\def\gevms{{\,\mbox{GeV}^{-2}}}

\def\({\left(} 
\def\){\right)}

\def\ie{\hbox{\it i.e.}}         
 
\def\eg{\hbox{\it e.g.}}

\def\pder#1#2{{\partial #1\over\partial #2}}

\def\beq{\begin{equation}} 
\def\eeq{\end{equation}} 
\def\bea{\begin{eqnarray}} 
\def\eea{\end{eqnarray}} 
\def\eq#1{{\mbox{Eq.\hspace{1mm}(\ref{#1})}}} 
 
\def\fig#1{{\mbox{Fig.\hspace{1mm}\ref{#1}}}}

\def\scrbox#1{\mbox{\scriptsize #1}}

\long\def\symbolfootnote[#1]#2{\begingroup%
    \def\thefootnote{\fnsymbol{footnote}}\footnote[#1]{#2}\endgroup} 
 
\def\epj#1#2#3{    {\it Eur.\ Phys.\ J.\ }       {\bf #1} (#2) #3}

\newcommand{\email}[1]{${\!}^{\scrbox{#1)}}$}

\newcommand{\chisquare}{\chi^2/\mbox{n.d.f.}}

\newcommand{\rv}{\vec{r}} 
 
\newcommand{\hrpar}{\hat{r}_{||}} 
\newcommand{\bv}{\vec{b}} 
\newcommand{\dsize}[1]{\vec{r}_{#1}} 
\newcommand{\sdsize}[1]{r^2_{#1}} 
\newcommand{\bdsize}[1]{\vec{b}-\half\vec{r}_{#1}} 
\newcommand{\dn}[1]{{\Delta N_y^{(#1)}}}

\relax 
\begin{document} 
\begin{titlepage} 
\noindent 
\begin{flushright} 
\parbox[t]{14em}{
\begin{tabular}{ll}
TAUP &   2757/2004\\ 
\end{tabular}
\\
{\tt  \today}
} 
\end{flushright} 
\vspace{1cm} 
\begin{center} 
  {\Large \bf 
Towards a New Global QCD Analysis: Solution to the Non-Linear
Equation at Arbitrary Impact Parameter
}
  \\[4ex] 
\begin{center}\large{ 
        E.~Gotsman   \email{1},  
        M.~Kozlov     \email{2},  
        E.~Levin     \email{3},  
        U.~Maor      \email{4} and  
        E.~Naftali   \email{5}} 
\end{center} 
 
\footnotetext{\email{1} gotsman@tau.ac.il }
\footnotetext{\email{2} kozlov@tau.ac.il} 
\footnotetext{\email{3}leving@tau.ac.il }
\footnotetext{\email{4} maor@tau.ac.il } 
\footnotetext{\email{5} erann@tau.ac.il } 
\vfill 
{\it  School of Physics and Astronomy}\\ 
{\it Raymond and Beverly Sackler Faculty of Exact Science}\\
{\it Tel Aviv University, Tel Aviv, 69978, ISRAEL}\\[4.5ex] 
\vfill

\end{center} 
~\,\, 
\vspace{1cm}

{\samepage {\large \bf Abstract:}}  
\vfill  
A  numerical solution is presented for the non-linear evolution equation 
that governs the dynamics of  high parton density QCD. It is shown that 
the 
solution falls off as $e^{-b/R}$ at large values of the impact parameter 
$b$. The 
power-like tail  of the amplitude appears  in impact parameter 
distributions  only after the  
inclusion  of dipoles of size larger than the target, a  
configuration for which the 
non-linear equation is not valid. The value, energy and impact parameter 
of the saturation scale $Q_s(y=\ln(1/x),b)$)  are calculated both  for 
fixed and 
running QCD coupling 
cases. It is shown that the solution exhibits    geometrical scaling 
behaviour.  
The radius of interaction increases as the rapidity  in 
accordance with the Froissart theorem.  The  
solution we obtain differs from  previous attempts, where an anzatz 
for $b$ behaviour was made. The solutions 
for running and fixed $\as$ differ. For running $\as$ we 
obtain a larger radius of interaction ( approximately twice as large), a
steeper rapidity  dependence, and a larger value of the saturation scale.

 
\end{titlepage} 
\section{Introduction} \label{sec:introduction} 
The main goal of this paper is to find a numerical solution
(including its impact parameter dependence)
 of the
 non-linear  evolution
equation \cite{BK}, that governs the dynamics of the dipole scattering 
amplitude in the saturation region \cite{GLR,MUQI,MV}. 

  Although, there are several schemes for finding  a 
numerical 
solution \cite{BR1,LL1,GBMS,WEI}, we still lack a reliable
method that will allow us to study the properties of the solution
of the non-linear evolution equation.
  For example, there is  the difficulty of
 understanding the large impact parameter ($b\equiv|\bv|$) behavior of the
 solution.  Our notation is such that the symbol ``$\rightarrow$'' 
represents a
 two dimensional vector, in the transverse plane.

The large $b$ behavior is strongly affected by the non-perturbative
 contributions \cite{KW}, since in pQCD the amplitude falls off
as a power of $b$.
  A decrease of this type
leads to a power-like growth of the interaction radius,
 which violates  the Froissart bound \cite{FROI}.  This difficulty
can be overcome  
 by  introducing non-perturbative corrections for large $b
\,\, (\,\,>\,\,1/m_\pi)$, where $m_\pi$ is the mass of the lightest hadron
 ($\pi$-meson). Two suggestions of how to include such non-perturbative
 corrections have been made : {\it(i)} to change the kernel of linear and
 non-linear equations \cite{KW}; and {\it (ii)} to accumulate all
 non-perturbative corrections in the initial conditions of the equation
 \cite{LRREV,FIIM,BKL}. The first attempt to solve the full non-linear
 equation including the impact parameter behavior, was made in
 Ref.~\cite{GBS}. The solution of Ref.~\cite{GBS}, shows that
to obtain the correct behavior at large
 values of $b$,
one has to alter the kernel of the equation.

 We examine this solution and show that the entire power-like
 decrease obtained in Ref.~\cite{GBS}, originates in the kinematic
 region where  the non-linear equation is not trustworthy.

In our search for the solution we are guided by:
(i) the known analytic 
expressions for the
 simplified cases \cite{LT,ML1,BKL}, and (ii)
   practical experience gained in
 previous attempts:
\begin{enumerate}
 \item \quad The non-linear evolution equation can be viewed as the sum of
 the semi-enhanced, `fan' diagrams (see \fig{fig:fan}a for such a
 diagram). However, we should restrict ourselves to summing such diagrams
 only for the scattering amplitude of a dipole  of transverse size
 ($|\rv|$), which is much smaller than the size of the target 
\cite{GLR,MUQI}
 ($|\rv|\ll R$, where $R$ is the target transverse 
size)\symbolfootnote[2]{For
 more detailed discussion of this key property of the non-linear equation see
 Refs. \cite{RY,LL}.}.  If the  target is smaller than  the
 projectile, we need to sum diagrams of the type shown in
 \fig{fig:fan}b.  Although, 
  the kernel of the non-linear equation does not depend explicitly
 on the size of the target, this quantity appears in the 
 solution by way of the region of 
 validity of the equation.

\item \quad In  the saturation region ($\rv\leq\,2/Q_s$, where $Q_s$ 
is
 the saturation scale) the solution displays geometrical scaling 
\cite{LT, BKL,
 GSC,MV}, namely, the scattering amplitude $\cal{A}$ is a function of
 only one variable
\beq \label{GSC} 
N\left(\rv,b;x \right)\,\,\equiv\,\,
\mbox{Im} A\left( \rv,\bv;x   \right)\,\,\,=
\,\,\,N\left(r^2 \,\,Q_s(x;b) \right). 
\eeq
We  therefore expect  that the entire $b$ dependence can be absorbed into
 the saturation scale $Q_s(x;b)$.

\item \quad We examine the behaviour of the scattering amplitude
$N\left(\rv,\bv;x\right)$, both for the case of the running QCD 
coupling constant, and for fixed $\alpha_{s}$.

\item \quad In the saturation region $N\left(\rv,\bv;x\right)< 1$,  
consequently
 the non-linear equation has a solution that satisfies $s$-channel
unitarity. On the other hand, the behavior of the total cross section:
\beq \label{TOTXS}
\sigma^{\mathrm{dipole}}(\rv,x)\,\,=\,\,2\,\,\int\,d^2\,b\,\,N\left(\rv,b_t;x
\right)\, 
\eeq 
depends critically on the large $b$ dependence of the amplitude, 
 and has to be studied separately.
 \item \quad The semiclassical solution \cite{BKL} to the non-linear
evolution equation,
 indicates that the behaviour at large $b$ stems  from
 the initial conditions.

\end{enumerate}  
 Our strategy is to assume that the size of the interacting dipole is
 smaller than that of the target. In the next section we describe the 
method
 of solution which we develop based on this strategy. In Section 3 we present
 our results. In particular, we check whether the geometrical scaling
 behavior in the form of \eq{GSC} occurs in our solution. In the 
Conclusions,
 we discuss the main results and the possibility of using them for  
 phenomenology.

\section{Method of the Solution} \label{sec:method} 

The nonlinear evolution equation\cite{BK} characterizes the low $x$ behavior
 of the parton densities, while taking into account hdQCD effects, so 
that 
 unitarity constraints are inherently obeyed.

We denote by $N$ the imaginary part of the interaction amplitude of a
 target with a parent dipole, of transverse size $\rv$ and two dipoles, of
 transverse sizes $\dsize{1}$ and $\dsize{2}\,$ ($\rv=\dsize{1}+\dsize{2}$),
 produced by $\rv$.  

The probability for the decay $\rv\rightarrow\dsize{1},\dsize{2}$ is
 given by the square of the wave function of the parent dipole, which,
 in a simplified form, can be written as $r^2/\sdsize{1}\sdsize{2}$.
 Each of the produced dipoles can interact with the target
 independently, with respective amplitudes 
 $N(y,\dsize{2};\bdsize{1})$ and $N(y,\dsize{1};\bdsize{2})$, where
 $y$ denotes the rapidity variable ($y=-\ln x$) and $\bv$ the impact
 parameter in the transverse plane.  Adding these contributions,
 clearly overestimates the dipole-nucleon interaction, since there
 is the probability that during the interaction, one dipole can be
 in the shadow of the other. This correction is given by an additional
negative quadratic
  term $-N(y,\dsize{2};\bdsize{1})N(y,\dsize{1};\bdsize{2})$.

\begin{figure}[t] 
\begin{center} 
\hspace{1.5cm}(a)\hspace{6cm}(b)\\
\includegraphics[width=5cm, bb=100 250 596 602]{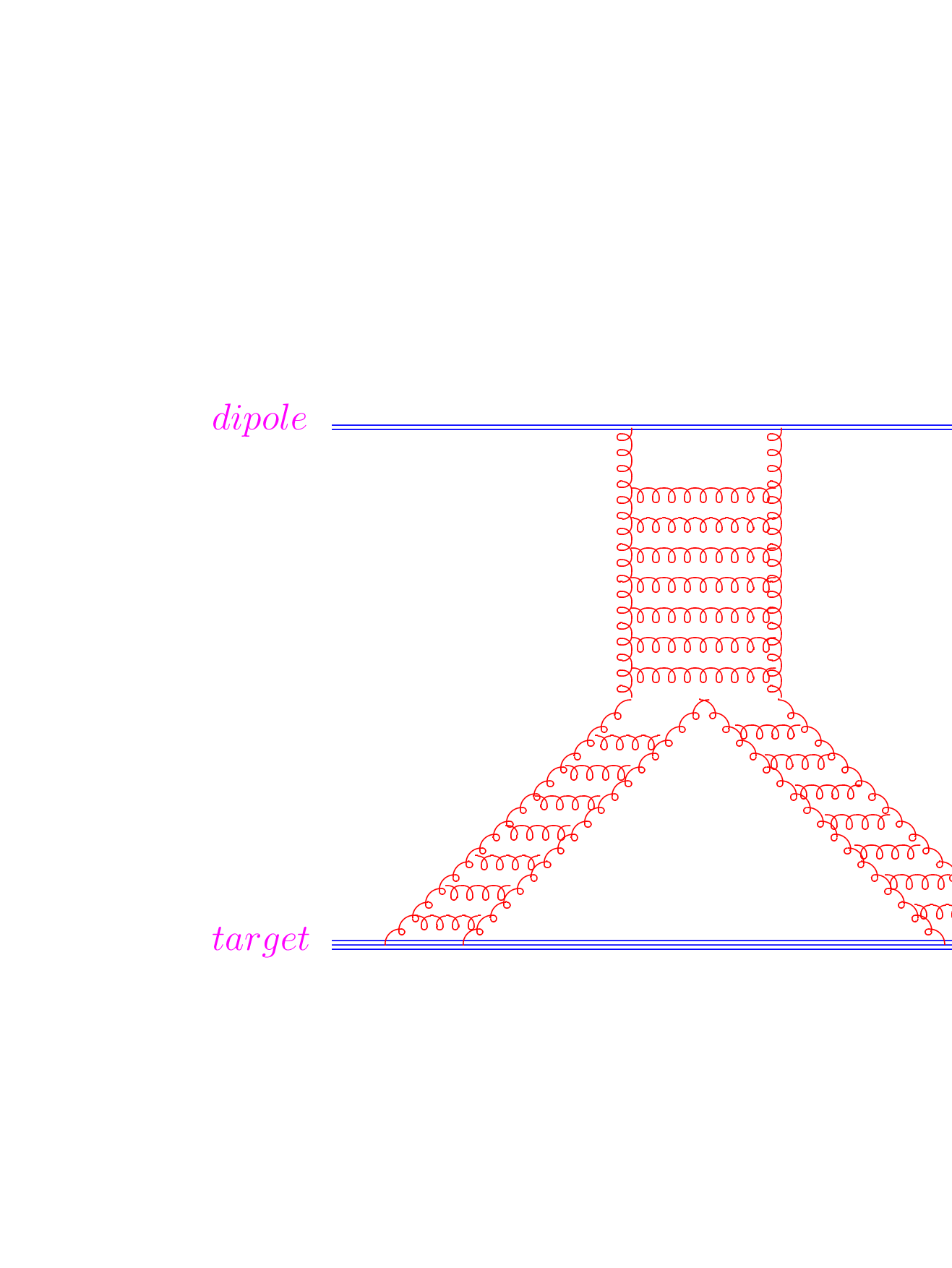}
\hspace{1cm}
\includegraphics[width=5cm, bb=100 250 596 602]{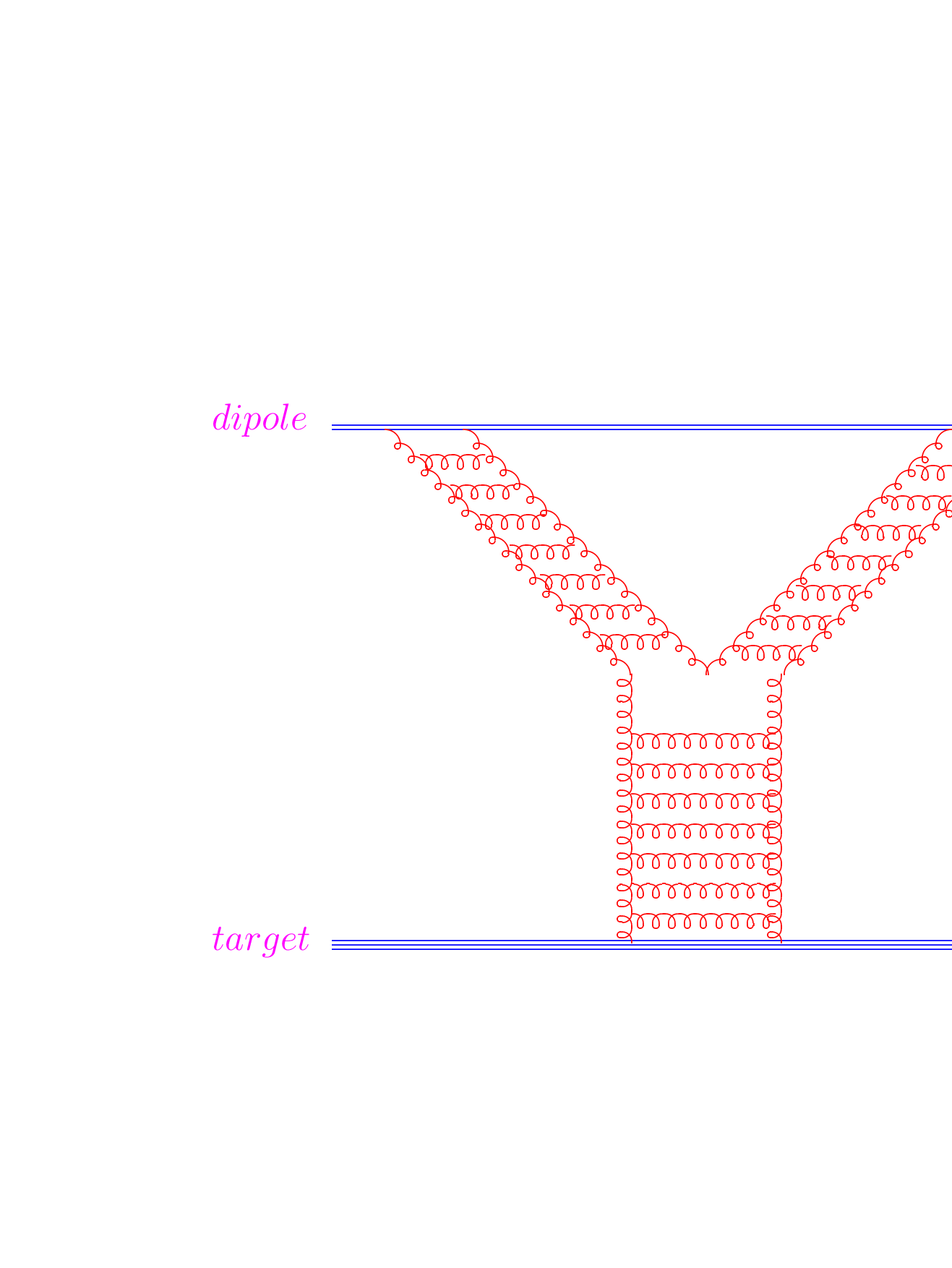}
\end{center} 
  \caption[]{\parbox[t]{0.80\textwidth}
{\small Multiple emissions of small $x$ gluons: (a) from the target, and (b)
  from the dipole.
}}
\label{fig:fan} 
\end{figure} 

In the procedure leading to the construction of the non-linear evolution
 equation, only fan diagrams of the type shown in \fig{fig:fan}a were taken
 into account, while fan diagrams having multiple emissions from the parent
 dipole (\fig{fig:fan}b) were omitted \cite{BK}. In other words, the
 equation does not include contributions for $|\dsize{1}|,|\dsize{2}| > 
R$,
 where $R$ is the transverse size of the target.  The solutions, that have
 been presented previously in Refs. \cite{BR1,LL1,GBMS,GBS}, included 
 contributions from both large and small size 
dipoles\symbolfootnote[3]{We would
 like to mention that the point of view presented in Ref. \cite{BR3} is
 very similar to ours.}.

Our approach is initially to exclude contributions from large dipoles, 
and
 then to estimate the accuracy of this proceedure by employing a second
 iteration.  The price we pay for this  approach is,  that
 our calculations are only valid for short distances,  compared to the 
size
 of the target.  Nevertheless, as we shall demonstrate later,
 contributions coming from large dipoles, which we consider as a
 correction, are relatively small.

Hence, we can write the evolution equation in a differential form as:
\begin{eqnarray} 
\nonumber \lefteqn{
N_y(y,{\rv};\bv)\equiv\pder{N(y,{\rv};\bv)}{y} = \frac{2C_F\as}{\pi^2}
\int d^2 r_2 \frac{r^2}{\sdsize{1}\sdsize{2}}
\,\theta(R-|\dsize{1}|)\,\theta(R-|\dsize{2}|)\, } \\ 
\label{eq:BK} & & \left[ 
2 N(y,\dsize{1};\bdsize{2}) - 
  N(y,\dsize{1};\bdsize{2})\,
  N(y,\dsize{2};\bdsize{1}) - 
  N(y,\rv;\bv) \right]\,.
\end{eqnarray} 
 The $\dsize{2}$ integration in (\ref{eq:BK}) is regulated by introducing an
 ultraviolet cutoff, which does not affect the physical quantities.

 Strictly speaking, the evolution equation is defined
 for fixed strong coupling, $\as$.  Since for running $\as$ it is
 difficult to define the running scale in terms of the variables of
 integration, $\dsize{1}$ and $\dsize{2}$, of the r.h.s of \eq{eq:BK}.  In the
 semiclassical approach, however, one sets $\as=\as (4/r^2)$, i.e.  one 
uses the
 size of the parent dipole, as the running scale of the coupling.  In the
 following, we shall attempt to compare  the results obtained with
 fixed coupling to those obtained with running coupling in the semiclassical
 approach.  Another approach would be to calculate the solution using
 $\as=\as((2\dsize{1}\dsize{2}/(\dsize{1}+\dsize{2}))^2)$, but such a
 calculation is beyond the scope of the present work. 

The strong coupling constant $\alpha_{s}$ is chosen to run according to
the prescription
$$ \alpha_{s}(r) \; = \; \frac{4 \pi}{\beta_{0}\;
\ln(4/(\Lambda^{2}_{QCD}r^{2}))} $$
where  $\Lambda_{QCD}$ = 0.2 GeV and
$\beta_{0}$ = (33 - 2$n_{f}$)/3. For the case of  fixed coupling constant
we take $\alpha_{s}$ = 0.2.

To obtain a solution to (\ref{eq:BK}) at arbitrary $b$, one needs to specify
 the initial conditions of $N$ at $y=y_0$, the rapidity from where we
 commence the evolution.  The initial condition is a function of both $\rv$
 and $\bv$ at $y=y_0$, while our goal is to obtain a  solution, which is
 a function of $\rv$ and $\bv$ for all $y \ge y_0$.  We are aware that during
 the evolution process, as $y$ increases, the role played by the initial
 input decreases, and the evolution becomes the dominant factor influencing
 the dynamics {\em per se}.

The experience  we have in solving \eq{eq:BK}
 (both with fixed and running
 $\as$) stems mostly from the solution of the equation, where shifts in
 $\vec{b}$ on the r.h.s.  of the equation, were neglected.  For such a
 simplified equation, the impact parameter is treated as a fixed parameter in
 the procedure.  For a solution of the type found in \cite{LL1}, once a trial
 solution was found at sufficiently large $y$, the curve approximating the
 solution was rescaled at fixed $b$ assuming geometrical scaling [see
 \eq{GSC}], so as to reproduce the initial conditions at $y=y_0$.  The
 reconstructed initial conditions were then used as input for the evolution,
 so that, the final solution almost completely decoupled from the initial
 input.  It was demonstrated in \cite{LL1} that the initial conditions
 obtained in this fashion, at very large distances, provide a smooth
 extrapolation of the Glauber formula to unity.

\begin{figure}[t] 
\begin{center} 
\includegraphics[width=14cm, bb=0 431 600 750]{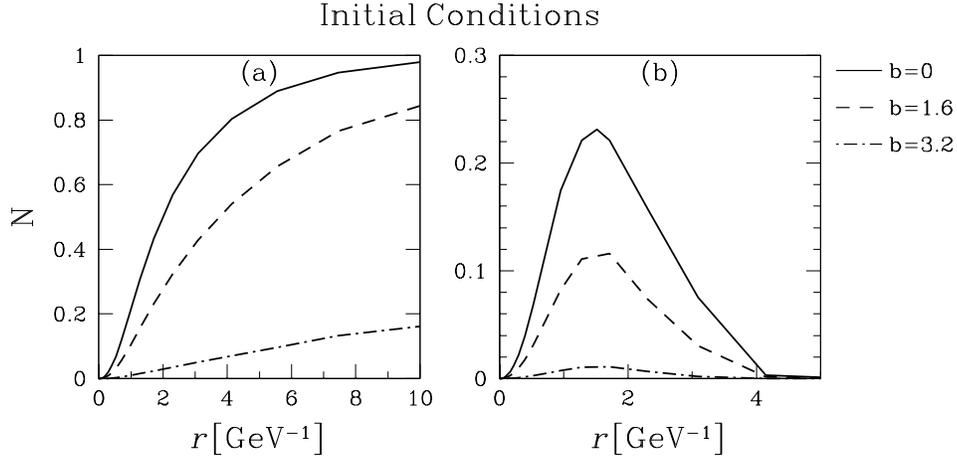}
\end{center} 
  \caption[]{\parbox[t]{0.80\textwidth}
{\small The solution of the non-linear evolution equation for different 
initial
conditions: (a) using \eq{Nb1} with a
Gaussian profile function; (b) using \eq{Nb2} with the same profile.
}}
\label{fig:initial} 
\end{figure}

In this paper we adopt the basic principle of this procedure, and use the
 rescaled solution of \cite{LL1} to construct the initial conditions. At
 first sight, it appears sufficient to follow \cite{LL1} by adopting the
 fixed $b$ solution as input,
 and assuming the following ansatz for the $b$-dependence:
\begin{equation} 
\label{Nb1}  
N(y_0,\rv; \bv)= 1\,-\,e^{-\kappa(y_0,\rv)\, S(\bv)/S(0)}\,,  
\end{equation} 
 where $S(\bv)$ is the profile function (\eg, a Gaussian or a dipole-like
 profile) and
\begin{equation} \label{kappa} 
\kappa(y_0,\rv)=-\ln\left(1\,-\,N(y_0,\rv;b=0)\right).
\end{equation} 
 However, in the Born approximation, at a finite mass of the gluon field, the
 scattering amplitude  for large size  parent
 dipoles, decreases already at $b=0$.
   To facilitate such  behavior, we  modify (\ref{Nb1}) 
as
 follows:
\begin{equation} 
\label{Nb2}  
N(y_0,\rv; \bv)= 1\,-\,e^{-\kappa(y_0,\rv)\, S(\sqrt{b^2+r^2})/S(0)}\,.
\end{equation} 
 Figs.~\ref{fig:initial}a-b show a comparison for the initial conditions
 of Eqs.~(\ref{Nb1}) and (\ref{Nb2}).  The replacement $b^2\longrightarrow
 b^2+r^2$, suppresses the initial conditions at large $r$, thereby 
smoothing the
 $\theta$ functions of (\ref{eq:BK}).

The most troublesome contribution in \eq{eq:BK} comes from the region of $r_1$
 and/or $r_2$ of the order of $2 b$. Indeed, in this region at large 
values
 of $b$, the r.h.s. of the master equation can be rewritten in the form
\beq \label{LBEST}
\Delta N \,\,=\,\,\frac{r^2}{b^4}\,\,\int d^2  \Delta b \,
\left(\,2\,N(y,b,\Delta b)\,-\,N^2(y,b,\Delta b)\,-\, N(y,r,b)\right)
\eeq
 One can see that $\Delta N\propto\,1/b^4\,$ at large $b$. This contribution
 has been discussed in Ref.~\cite{KW,GBS}. We claim that this 
contribution
 originates from the interaction of the dipoles of size much larger than 
the
 size of the target, and that it cannot be evaluated using \eq{eq:BK}.

Apart from  rapidity, the function $N$ depends on the dipole's degrees of
 freedom, which are the transverse separation of the parent dipole, $\rv$,
 and the transverse impact parameter, $\bv$.  We therefore, define the
 following variables, through which we express the dipole-nucleon amplitude:
 $r\equiv|\rv|$, $b\equiv|\bv|$, and $\hrpar\equiv\bv\cdot\rv/(b r)$. In the
 following, we shall concentrate on the integrated 
 quantity $N(y,r;b)\equiv\int d\hrpar N(y,r;b;\hrpar)$.

Once the initial conditions were established, 
we applied the following numerical procedure,
 which has 
 four stages: 

\begin{enumerate}
 \item {\bf First iteration:} we denote by $y_i$ a particular rapidity, at
 which the solution and its derivative are known for all values of $r\,, b$
 and $\hrpar$.  The solution at $y_{i+1}\equiv y_i+h(y)$ was constructed as a
 matrix, in which the matrix elements correspond to (fixed $y$) solutions at
 different $r$ and $b$, integrated over $\hrpar$:
\begin{equation}\label{eq:solution1}
N^{(1)}(r,y_{i+1};b)=N^{(1)}(r,y_{i};b)+h(y_i)\int d\hrpar 
\left. N_y^{(1)}(y,r;b;\hrpar)\right|_{y=y_i}\,,
\end{equation}
 where $N_y^{(1)}$ is given by the R.H.S of (\ref{eq:BK}), and $h(y_i)$ 
is a
 variable step size in the rapidity space. The first iteration included
 successive constructions of such matrices over a range of about 10 units of
 rapidity, starting from $y_0\approx 4.6$.  The value of $h(y_i)$ was chosen
 dynamically to ensure that the maximal error over the $i$th matrix is small
 (we set our accuracy condition to $10^{-3}$). We used the Euler two-step
 procedure\symbolfootnote[4]{In the Euler two-step procedure, one obtains a solution
 along the path $y_i\rightarrow y_i+h$ and an additional solution along the
 path $y_i\rightarrow y_i+h/2\rightarrow y_i+h$. The difference per unit step
 between the solutions is proportional to $h(y)\dd^2N/\dd y^2$, and the
 convergence criterion is that the maximal difference over the matrix is
 small.}
 to select $h(y)$, we  found that the optimal step size may
 vary from $h\approx 0.01$ for small rapidities to $h\approx 0.2$ for 
large
 rapidities.  The CPU time for the entire iteration was about $20$ hours for
 $30\times 30$ matrices.

Our main claim is that the first iteration as described above, provides a
 reliable solution to the  master equation.

\item {\bf First Correction:} in order to check the accuracy of the first
 iteration, and to estimate the importance of long distance 
contributions,
 we substituted $N^{(1)}$ back into the non-linear evolution equation, 
and
 recalculated $N_y$ in the kinematical region which was omitted from
 (\ref{eq:BK}), namely for $|\dsize{1}|,\,|\dsize{2}|>R$.  In other words, we
 calculated the contribution given by \eq{LBEST}.  We consider the result 
of
 this calculation  an estimate of the correction to $N_y^{(1)}$, and 
denote
 it by $\dn{1}$.

\item {\bf Second iteration:} to estimate the influence of the neglected
terms on the behavior of the solution in the saturation region, we
calculated the second iteration.  Namely, we used the calculated $\dn{1}$ (see
\eq{LBEST}) as a second iteration, by adding the correction to the r.h.s 
of
(\ref{eq:solution1}):
\begin{equation}\label{eq:solution2}
N^{(2)}(y_{i+1},r;b)=N^{(2)}(y_i,r;b)+h(y_i)\left(
\dn{1}(y_i,r;b) + \int d\hrpar
\left. N_y^{(2)}(y,r;b;\hrpar)\right|_{y=y_i}\right)\,.
\end{equation}

\item {\bf Second correction:} finally, we calculated $\dn{2}$, the
correction to the second iteration, by performing the integration over long
distances as before.
\end{enumerate}

This procedure gives us full control both on the accuracy of the 
solution,
and on the region of its  applicability.  
\begin{figure}[t] 
\begin{center} 
\includegraphics[width=8cm]{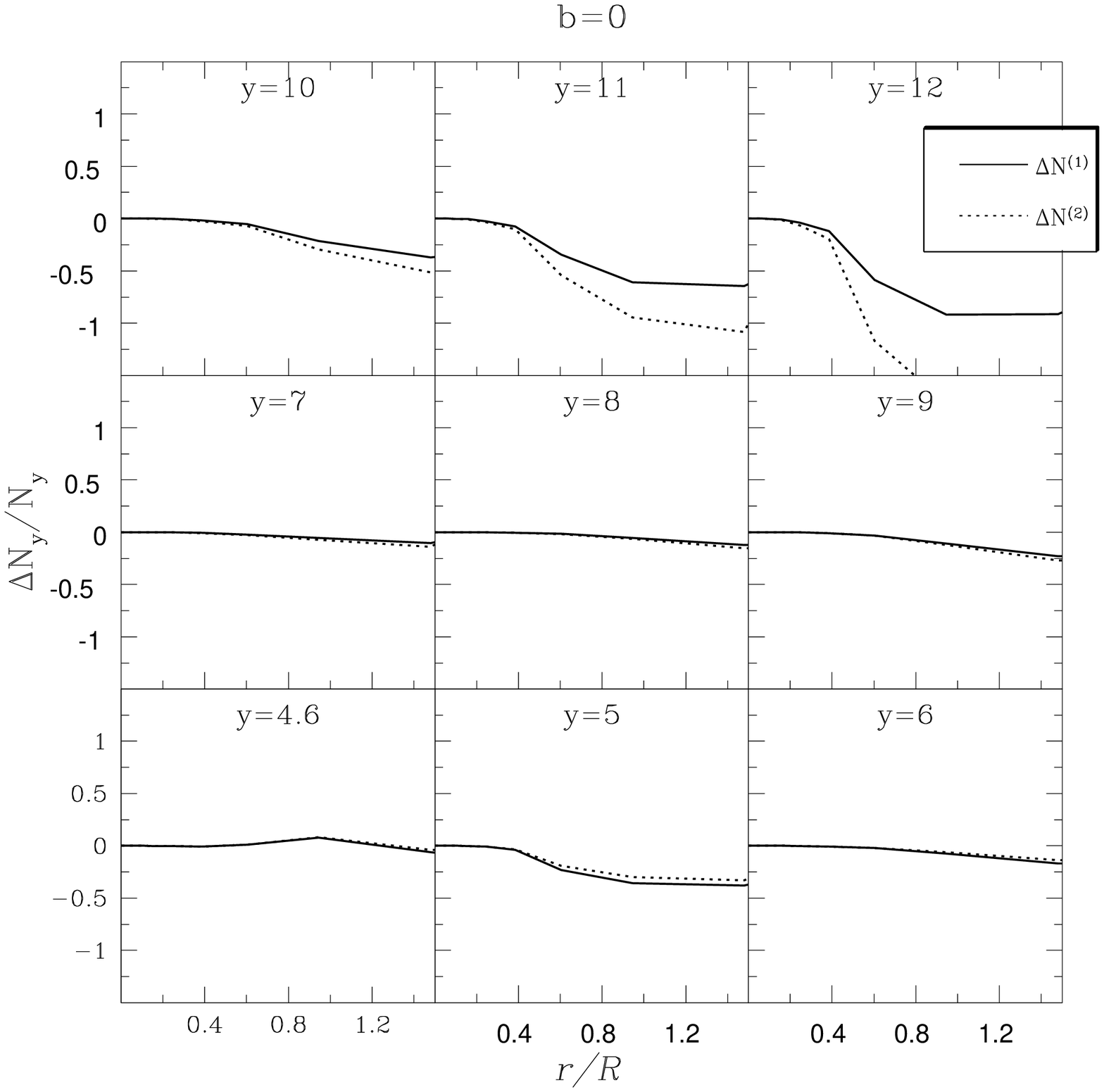} \includegraphics[width=8cm]{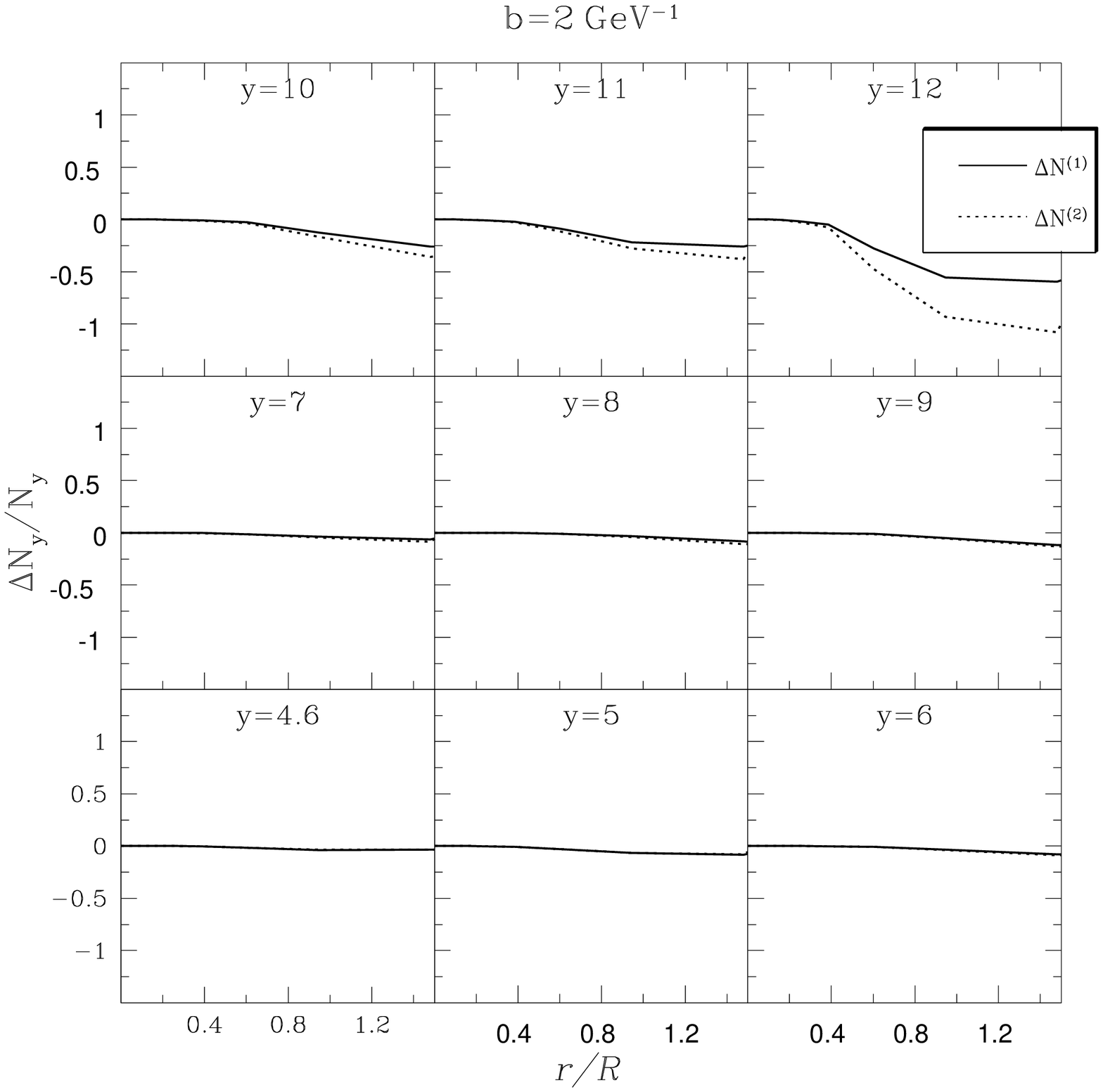}
\end{center} 
  \caption[]{\parbox[t]{0.80\textwidth}
{\small 
The corrections to the first and the second iterations for $b=0$ and
$b=2\,\mbox{GeV}^{-1}$ .
}}
\label{fig:dn0} 
\end{figure} 
 The corrections of the first and second iterations are shown in
 \fig{fig:dn0}, for $b=0$ and $b=2\,\mbox{GeV}^{-1}$, as a function of the
 ratio $r/R$. For not too large rapidities ($y<10$), the corrections are of
 order 10\percent, even when the transverse size of the dipole $r$, is
 comparable to the transverse size of the target, $R$.  For larger values of
 $y$, the corrections are  small only  for short distances.  The second
 iteration is important at large rapidities where $|\Delta
 N_y^{(2)}|<|\dn{1}|$. On comparing $\Delta N$ for both iterations, we 
see that we can only
 trust our solution  for $r/R \,<\,0.4$ at large values 
of
 rapidity ($y > 10$), while for small values of $y$  the  
accuracy of our solution is satisfactory,
 even for values of $r/R \approx 1.5$. This observation is most 
encouraging as we would like
 to use the solution to describe   experimental data, 
which are
 mostly concentrated  at $y < 10$.

\section{Evolution Results}
\begin{boldmath}
\subsection{$r$, $y$ and $b$  dependence:}
\end{boldmath}

\begin{figure}[t] 
\begin{center} 
\includegraphics[width=14cm, bb=0 480 580 710]{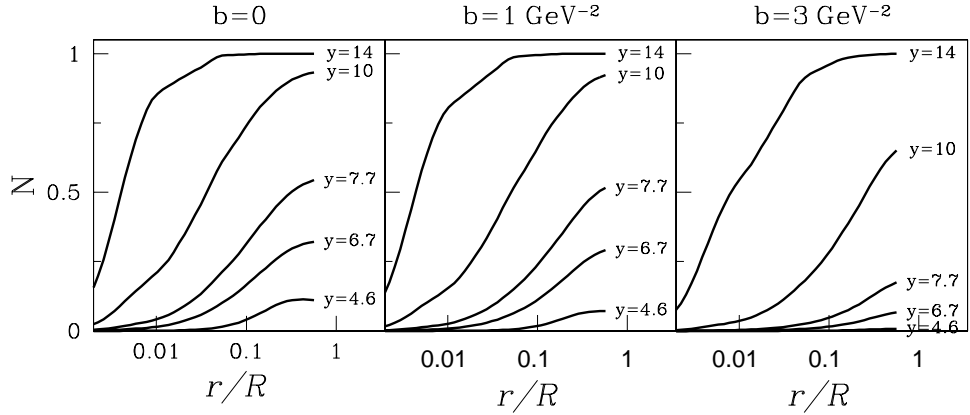}
\end{center} 
  \caption[]{\parbox[t]{0.80\textwidth}
{\small 
The evolution of $N$ as a function of $r/R$, for fixed $b$.
}}
\label{fig:rdep} 
\end{figure}

\begin{figure}[t] 
\begin{center} 
\includegraphics[width=10cm]{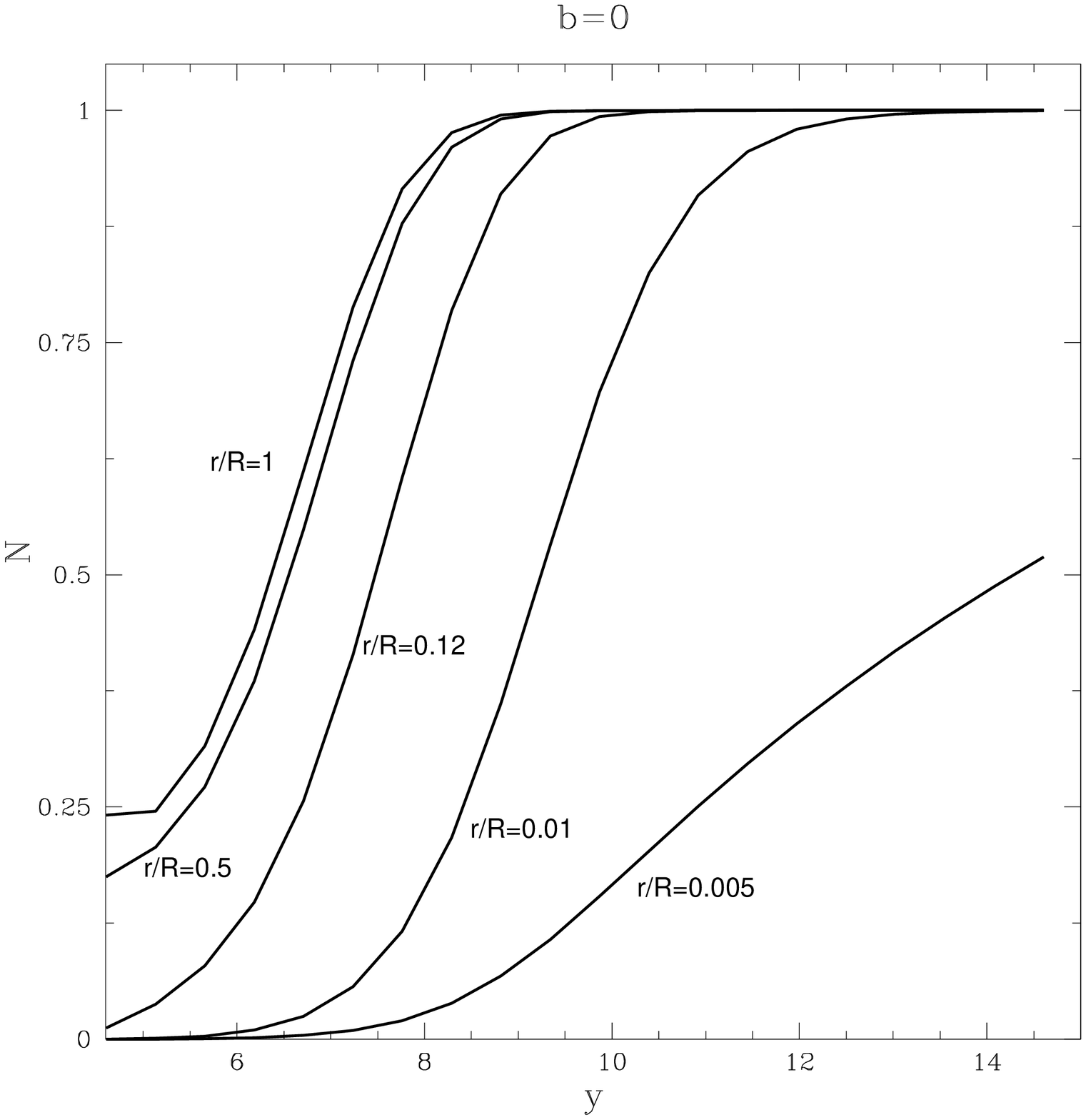}
\end{center} 
  \caption[]{\parbox[t]{0.80\textwidth}
{\small 
The energy dependence of $N$ at $b=0$, for different values of $r/R$.
}}
\label{fig:ydep} 
\end{figure}

\fig{fig:rdep} shows the evolution process from $y=4.6$ to $y=14$ at
 $b=0,\,1,\,3\,\gevm$ as a function of $r/R$, and \fig{fig:ydep} shows the
 energy dependence of the amplitude at $b=0$, for different values of 
$r/R$.
 We have chosen a target size of $R^2=3.1\gevms$ so as to compare
  with \cite{LL1}, where the fixed $b$ solution was successfully
 fitted to the HERA experimental $F_2$ data, using $R^2$ as a 
 parameter.
In the next section, we  discuss in more detail the differences and 
similarities
 between the present work and the solutions of \cite{LL1,GBS}.

Referring to \fig{fig:rdep}, we see that at the beginning of the evolution
 process, the amplitude  increases mildly with the dipole size, and 
does not exhibit
  saturation in the region of $r<R$.  As $y$ increases, the rise of $N$
 becomes steeper.  This rise is tamed when the impact parameter becomes
 large.

 In our approach, only the kinematical region $r/R\lsim 1$ can be
 explored accurately , although, there are some other kinematical
 regions where the corrections $\dn{1}$ and $\dn{2}$ are relatively 
small.
At long distances
 there  remain considerable theoretical uncertainties.

\begin{figure}[t] 
\begin{center} 
\includegraphics[width=12cm, bb= 20 150 570 700]{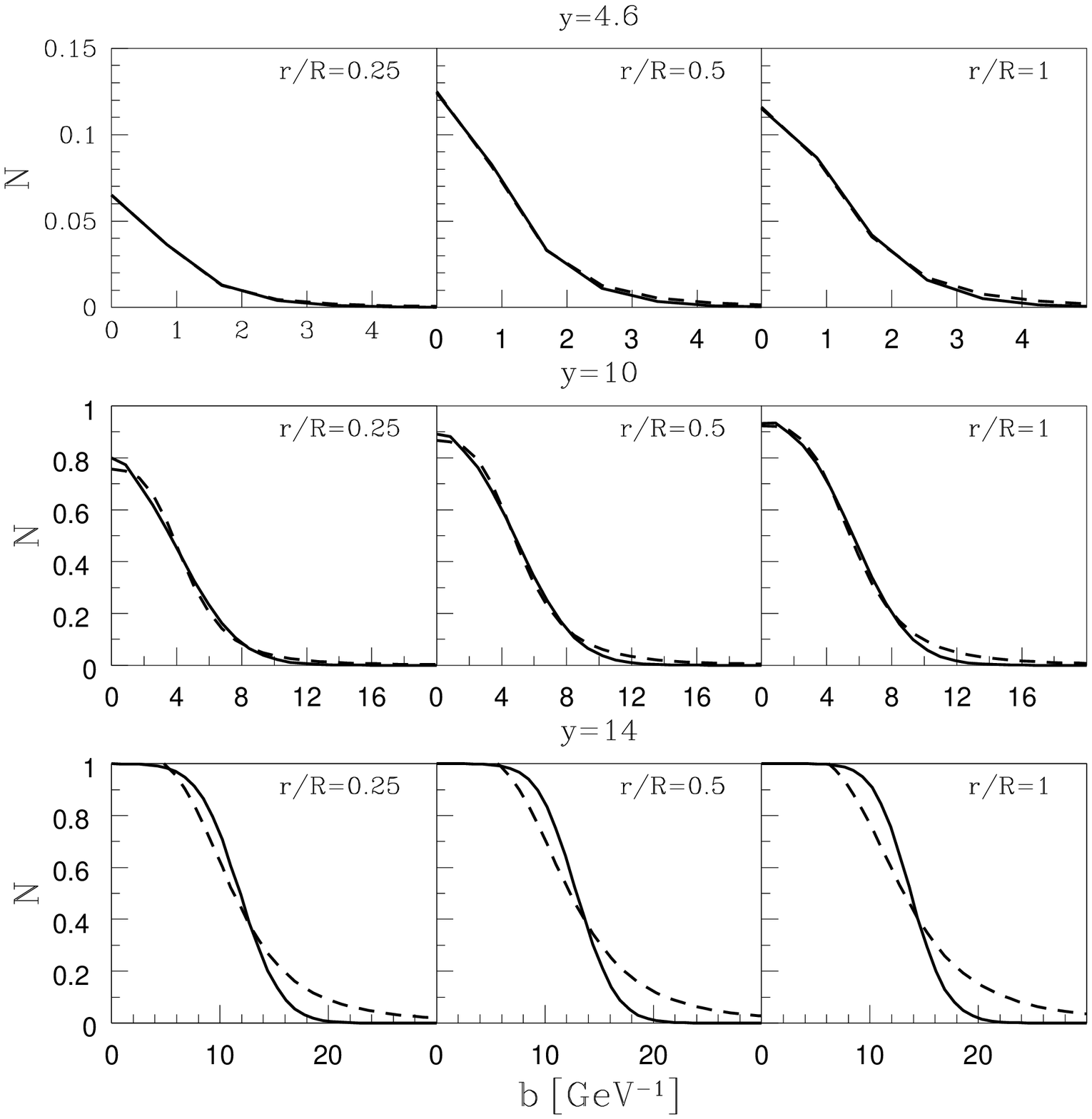}
\end{center} 
  \caption[]{\parbox[t]{0.80\textwidth}
{\small 
The impact parameter dependence of $N$ for $r/R$=1. The dashed line
corresponds to the $b^{-4}$ parameterization of \eq{eq:bpar}.
}}
\label{fig:bdep} 
\end{figure}

\begin{figure}
\begin{center}
\includegraphics[width=12cm, bb= 20 550 570 700]{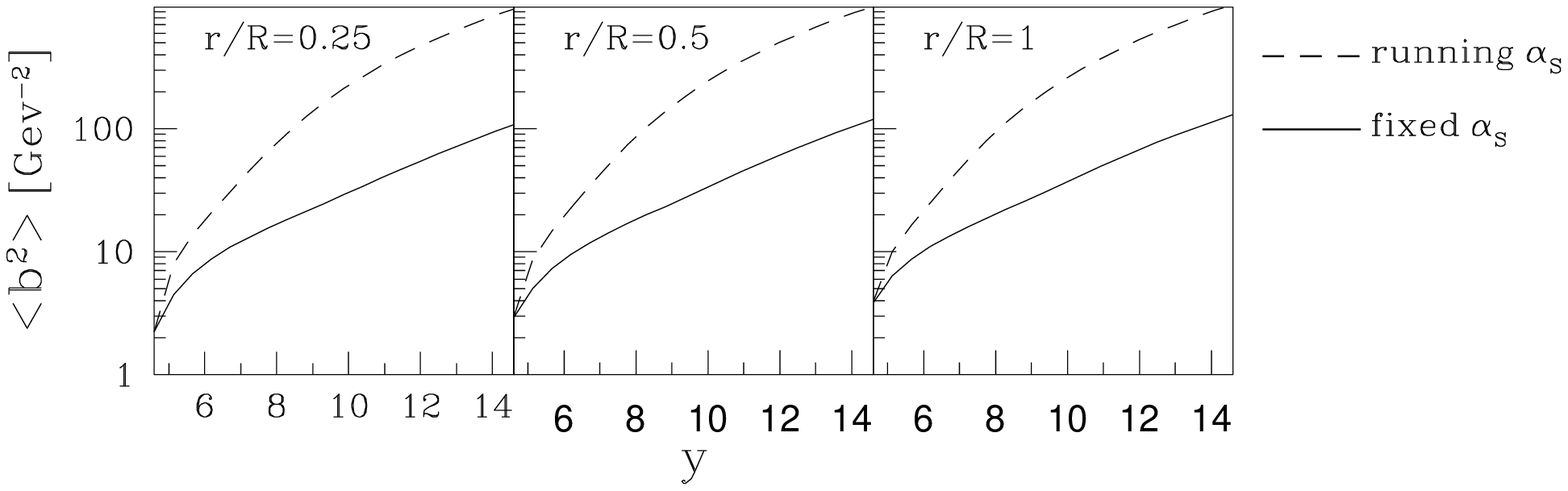}
\end{center}
  \caption[]{\parbox[t]{0.80\textwidth}
{\small Energy behavior of the average $b$ at different values of the
  dipole size for fixed (solid) and running (dashed) QCD coupling. 
}}
\label{fig:bav}
\end{figure}

\fig{fig:bav} shows the $b$-dependence of our solution for $r/R=1$. It 
was
 argued in \cite{GBS}, that  for small rapidities the evolution 
process
 starts to generate a power behavior of $b^{-3.6}$, which becomes even 
milder
 at higher rapidities.  The change in  behavior is due to large 
dipole
 configurations, which has been discussed above.  To illustrate the 
behavior
 of $N$ as a function of the impact parameter, we have parameterized our
 solution as:
 \begin{equation}\label{eq:bpar}
 N(y,r,b)=\frac{f(r,y)S(b)}{1+g(r,y)S(b)}\,,
\end{equation}
 where $S(b)=(R^2+b^2)^{-2}$ and $f$ and $g$ are fitted
 functions. \eq{eq:bpar} is presented in \fig{fig:bdep} as a dashed line.  In
 our approach, since we exclude large dipoles from the evolution, the
 amplitude at large $y$ deceases with $b$ faster than $b^{-4}$.
As was pointed out in Ref.\cite{KW} the power-like decrease leads to
 power-like growth of the average $b^2$ as a function of energy. However,
 \fig{fig:bav} 
 shows that at large values of the dipole size there is only a slow increase,
 (if at all).  Indeed, the simple formula $ \langle b^2 \rangle = a + b\,y + cy^2 $ 
provides
 a good description of the enegy behaviour of the radius of interaction, for
 different values of $r$.  (see \fig{fig:bav}). 

\begin{figure}[t] 
\begin{center} 
\includegraphics[width=12cm]{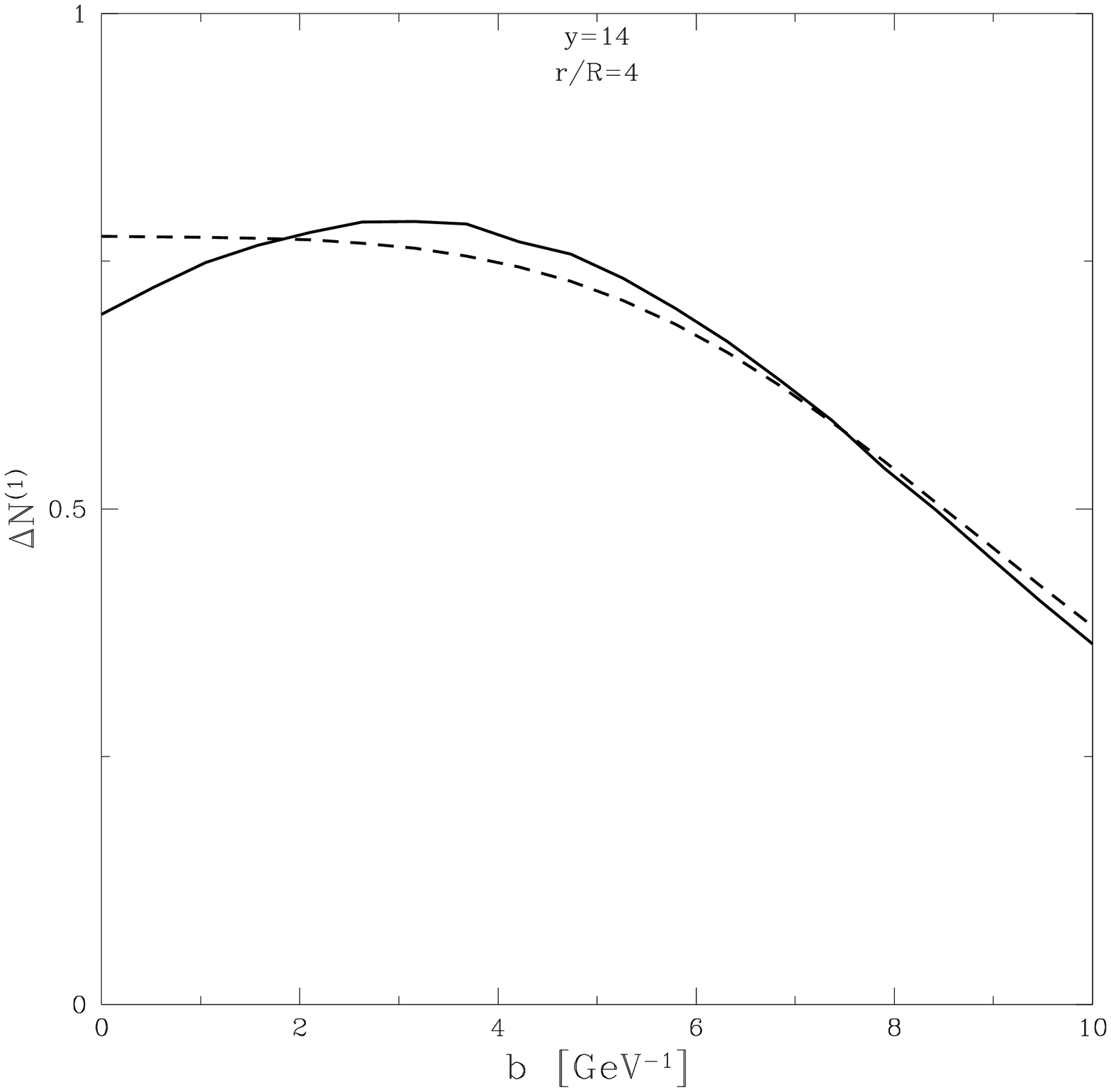}
\end{center} 
  \caption[]{\parbox[t]{0.80\textwidth}
{\small 
The impact parameter dependence of $\Delta N$. The dashed line corresponds to
the $b^{-4}$ parameterization of \eq{eq:bpar}, employed for $\Delta N^{(1)}$.
}}
\label{fig:bdepdn} 
\end{figure}

We have also used Eq.(\ref{eq:bpar}) to parameterize the correction $\Delta
 N^{(1)}\equiv\int\dn{1}dy$.  In \fig{fig:bdepdn} we show $\Delta N^{(1)}$
 and its $b^{-4}$ parameterization for large $r/R=4$ and large $y=14$, we
 stress that the power tail originates from the correction to the first
 iteration.

\begin{figure}[t]
\begin{center}
\epsfig{file= 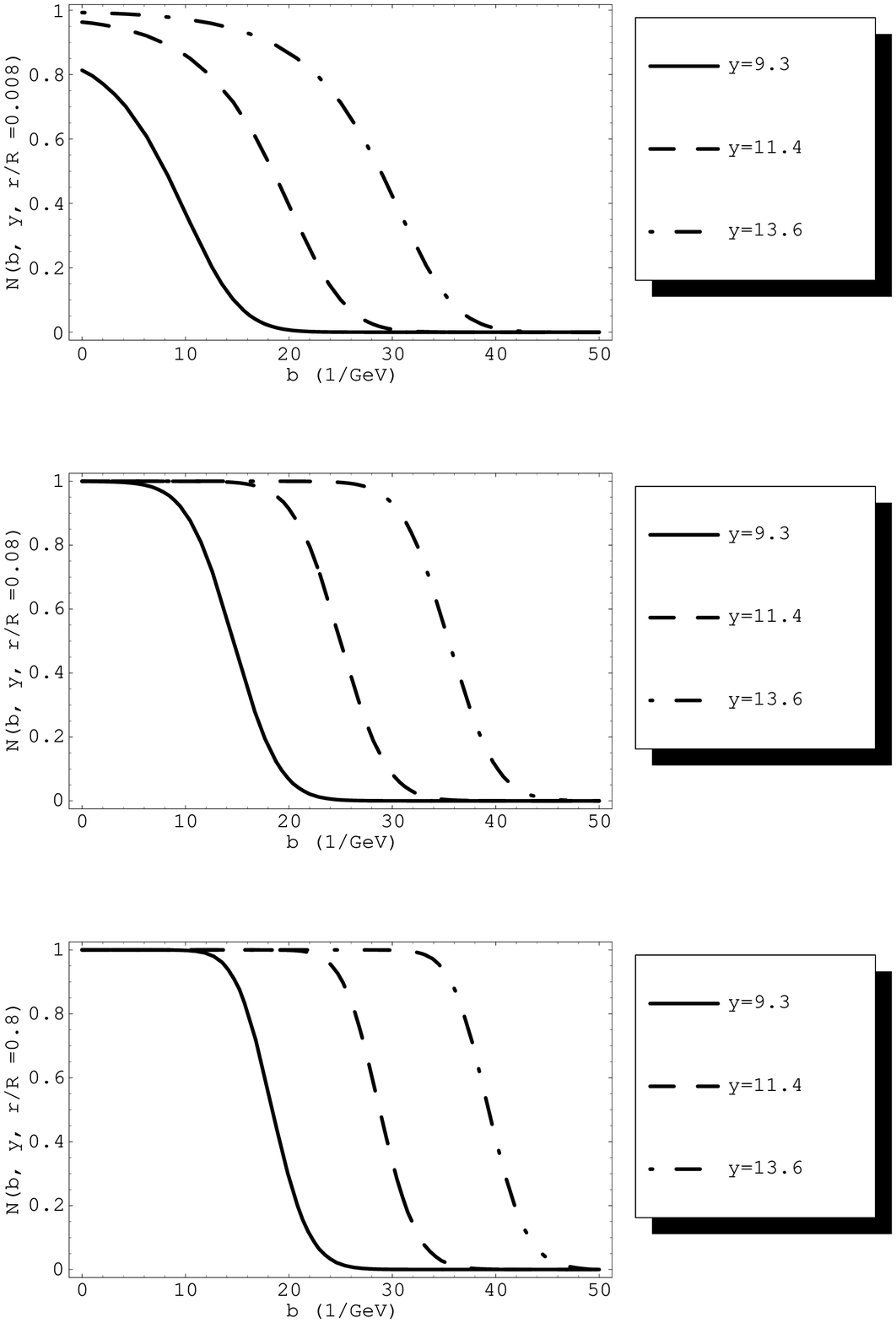, width=140mm,height=180mm}
\end{center}
\caption{Our solution for the scattering amplitude $N$ as function of the
impactparameter $b$, at different energies and at different values of $r$
 for running QCD coupling.}
\label{fig:Nbrun}
\end{figure}

\begin{figure}[t]
\begin{center}
\epsfig{file= 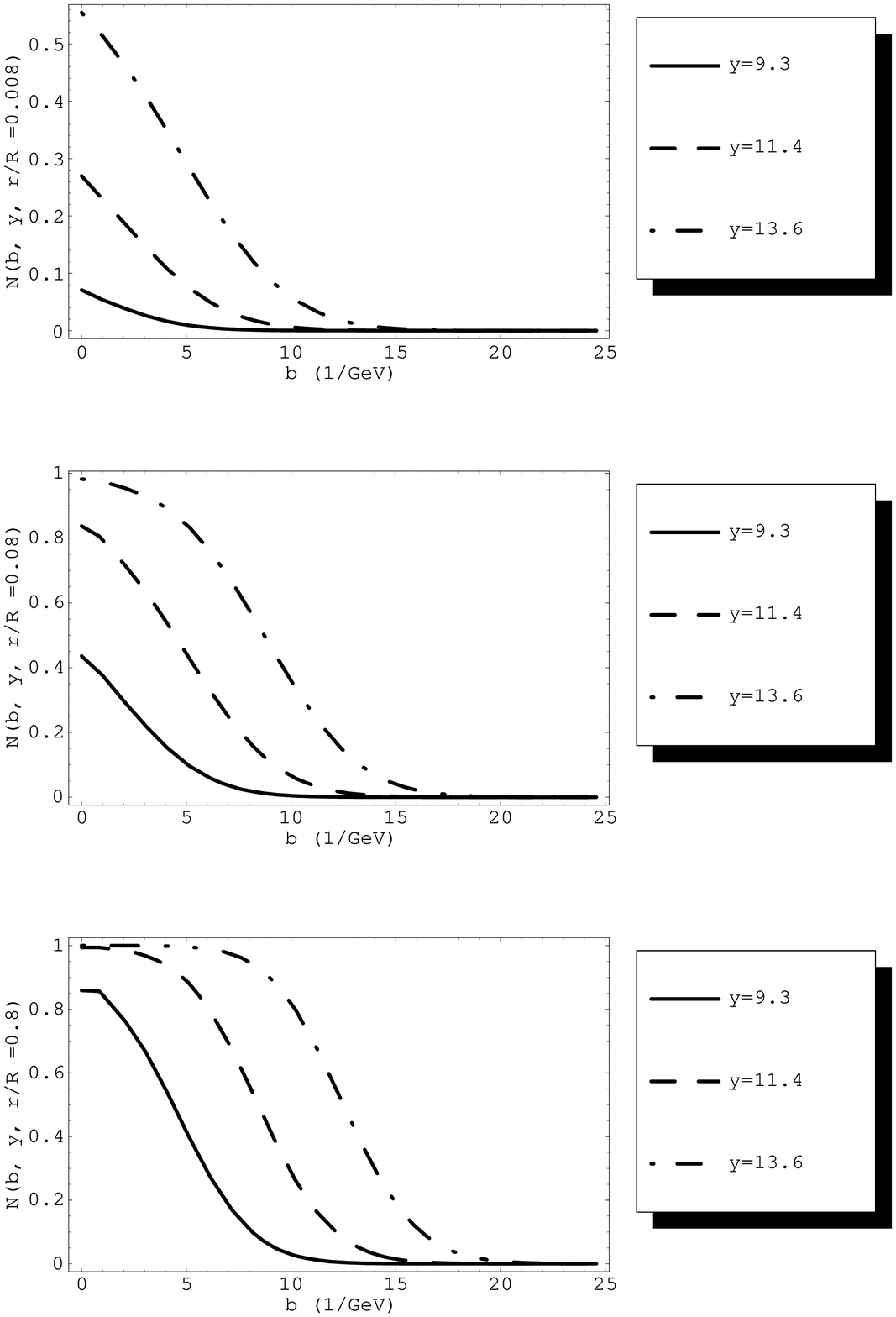, width=140mm, height=180mm}
\end{center}
\caption{Our solution for the scattering amplitude $N$ as function of 
the
impact parameter $b$, at different energies and at different values of $r$
 for fixed QCD coupling.}
\label{fig:Nbfix}
\end{figure}

\begin{figure}[t]
\begin{tabular}{c c  c}
Fixed $\as$ & & Running $\as$ \\
 ~ & ~ &\\
\epsfig{file= 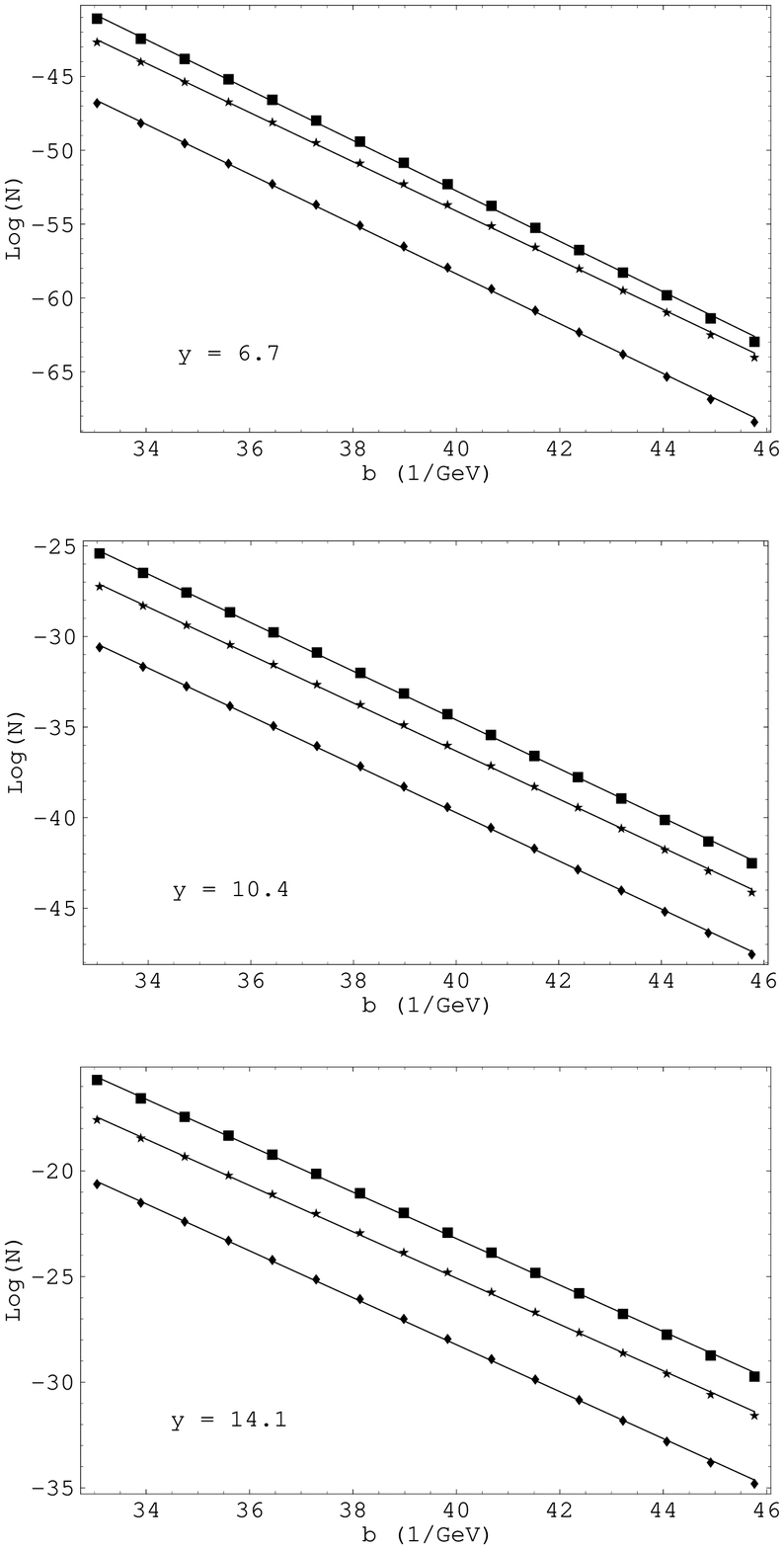, width=60mm,height=145mm} &
\epsfig{file= 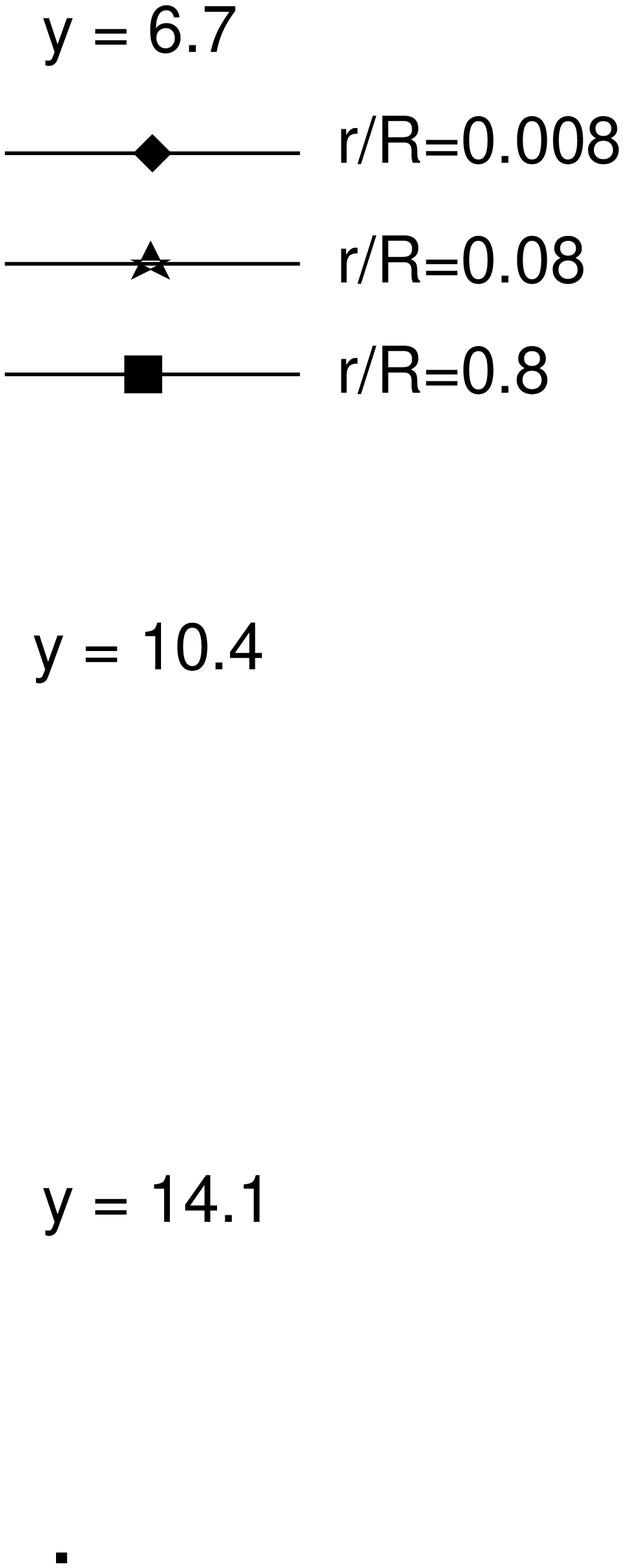, width=45mm,height=145mm} &
\epsfig{file= 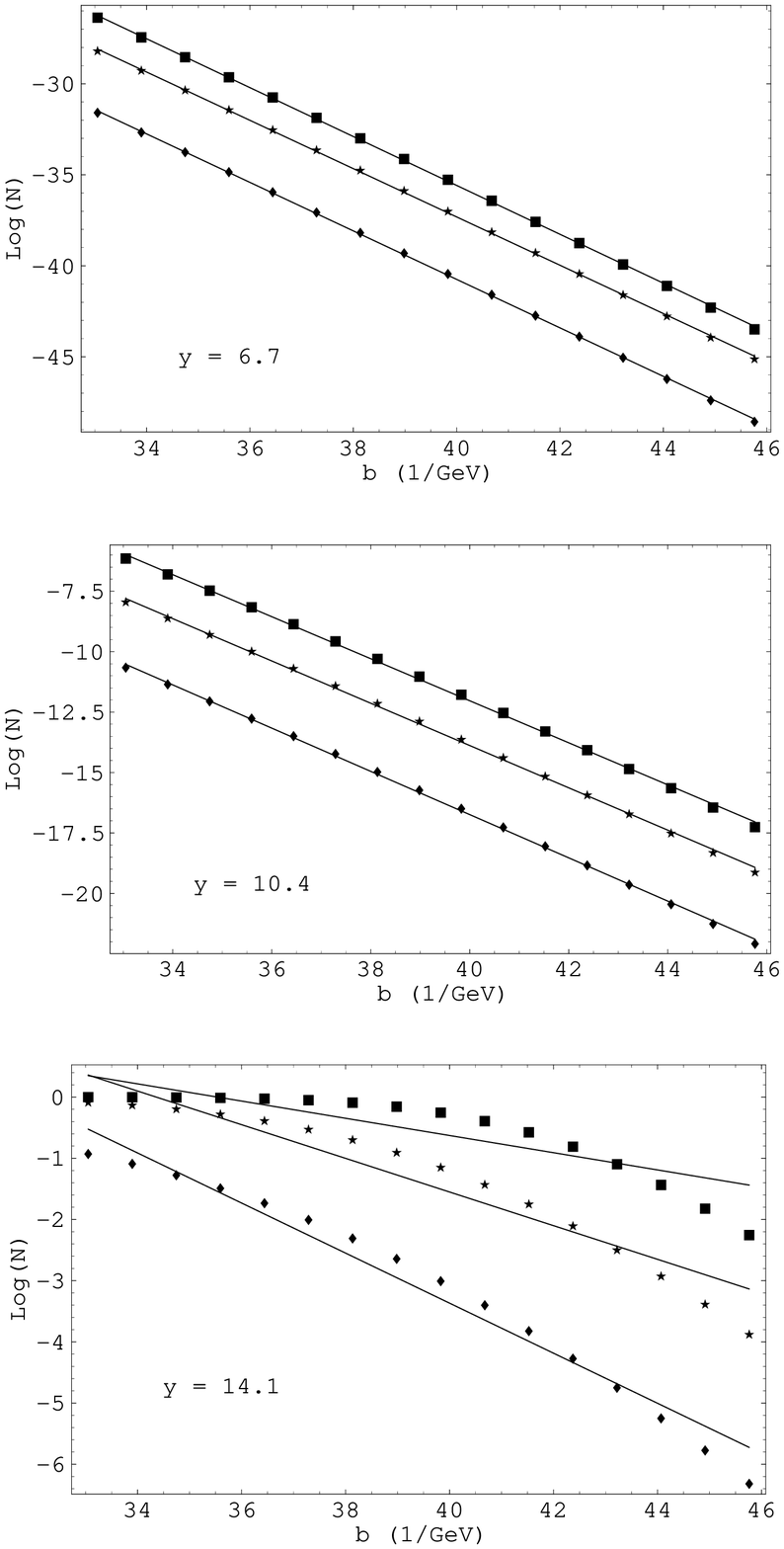, width=60mm, height=145mm}
\end{tabular}
\caption{The large impact parameter behaviour of the scattering
amplitude $N$ for fixed and running  $\as$.}
\label{fig:Nbfixtail}
\end{figure}

 \fig{fig:Nbrun} and \fig{fig:Nbfix} demontstrate the essential difference between 
running and fixed QCD coupling cases. One sees that for running QCD 
coupling the amplitude has a steeper growth with energy, reaching the 
saturation 
boundary for smaller $y$ than in the case of fixed $\as$. On the other 
hand, the behaviour of the tail in the impact parameter distribution 
looks 
very similar in both cases (see \fig{fig:Nbfixtail}). It should be 
stressed that the
scattering amplitude falls off exponentially at 
large $b$ ( $N \,\propto e^{ - m\,b}$ with
 $m$ = 1 to  2 GeV, see \fig{fig:Nbfixtail} ). \fig{fig:bdepdn} and 
\fig{fig:Nbfixtail} show 
that our solution has no power-like tails in the
 $b$ distribution, and therefore, supports our hypothesis 
that the power-like large $b$ decrease, stems from the
 dipole configuration which cannot be treated 
within the non-linear Balitsky - Kovchegov equation.

\subsection{Saturation scale}
\begin{figure}[t]
\begin{tabular}{c c}
Fixed $\as$ &Running $\as$ \\
~ & ~\\
\epsfig{file= 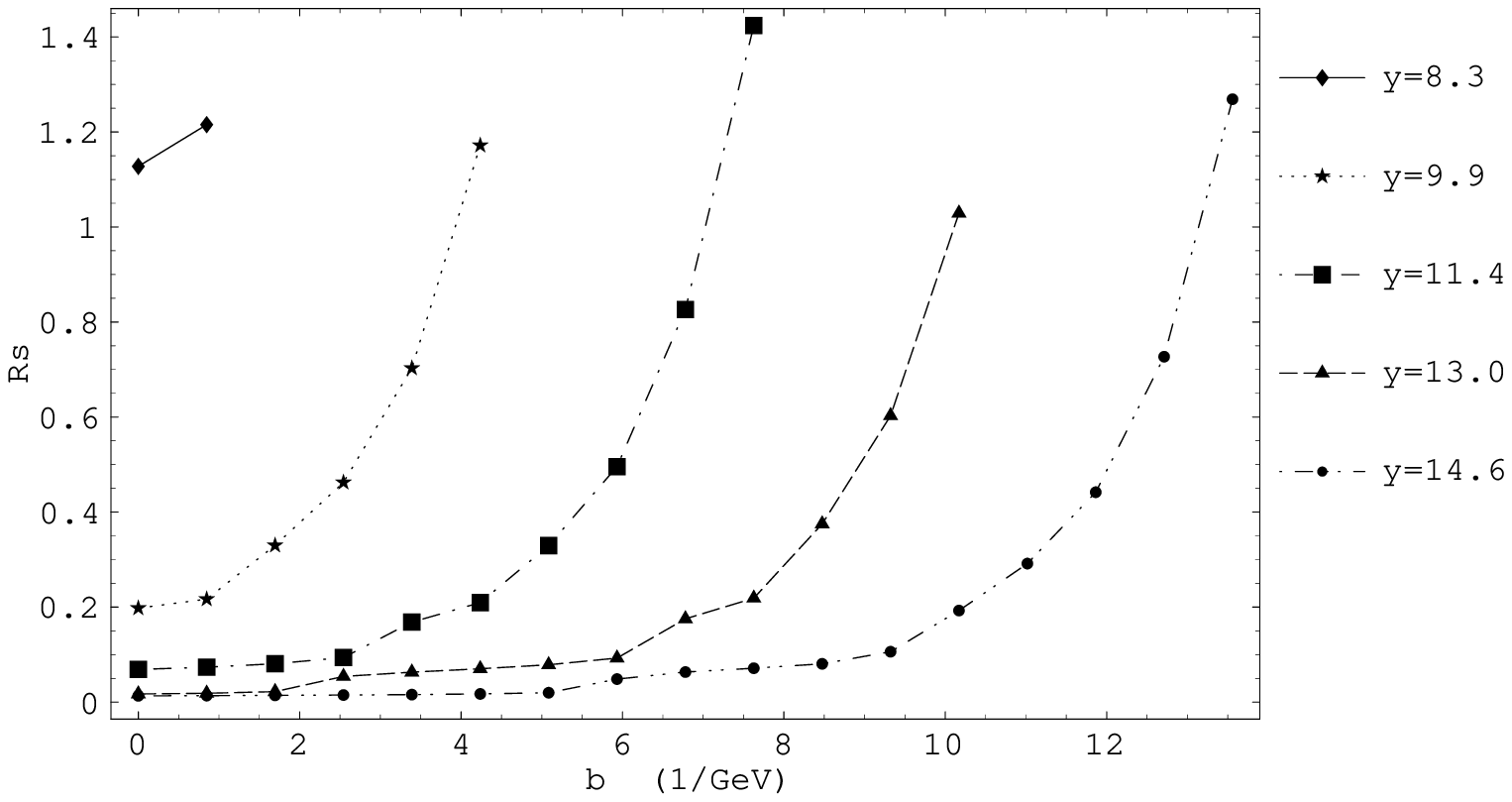, width=80mm} &
\epsfig{file= 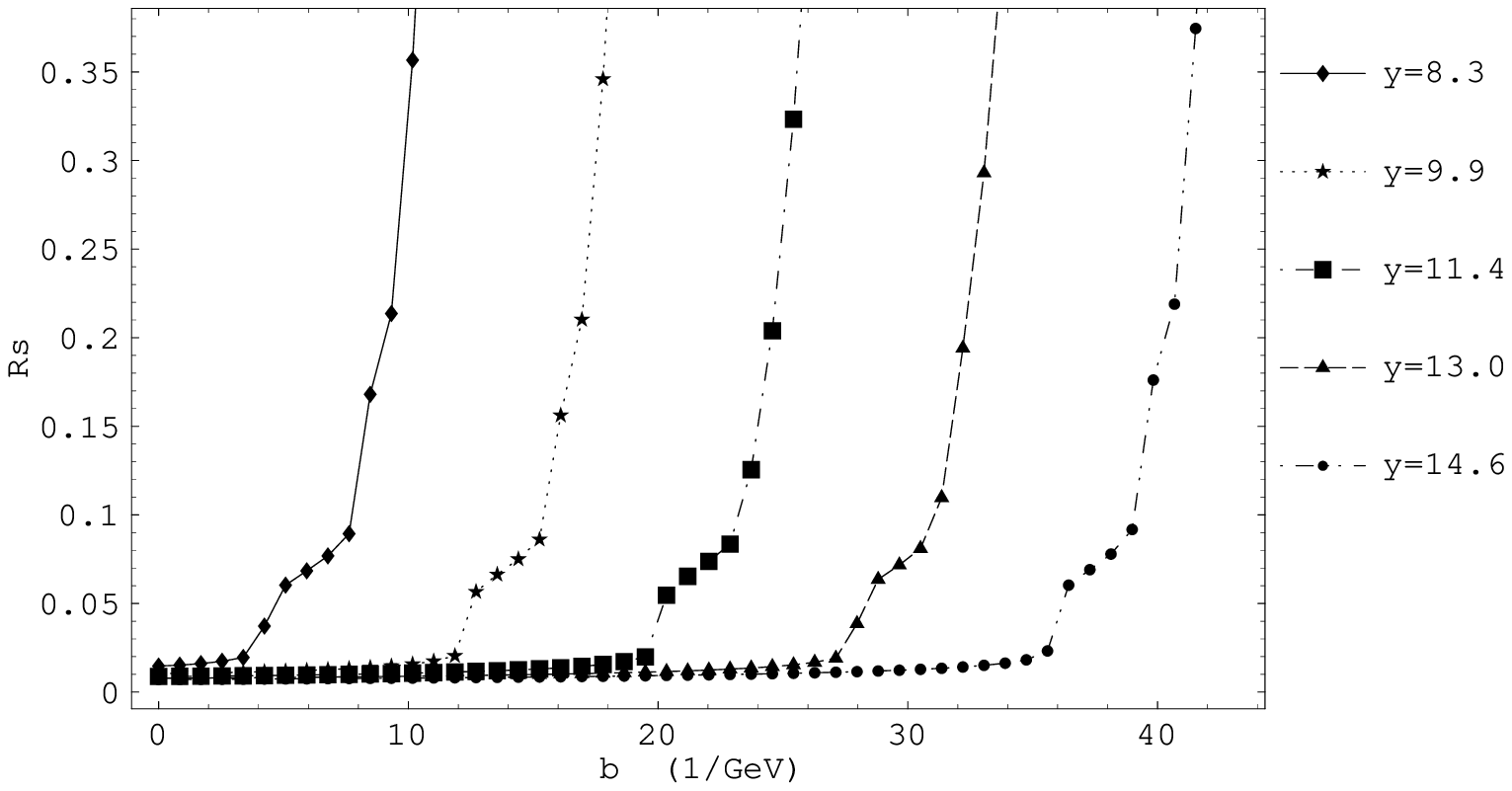, width=80mm}\\
 ~ & ~\\
\epsfig{file= 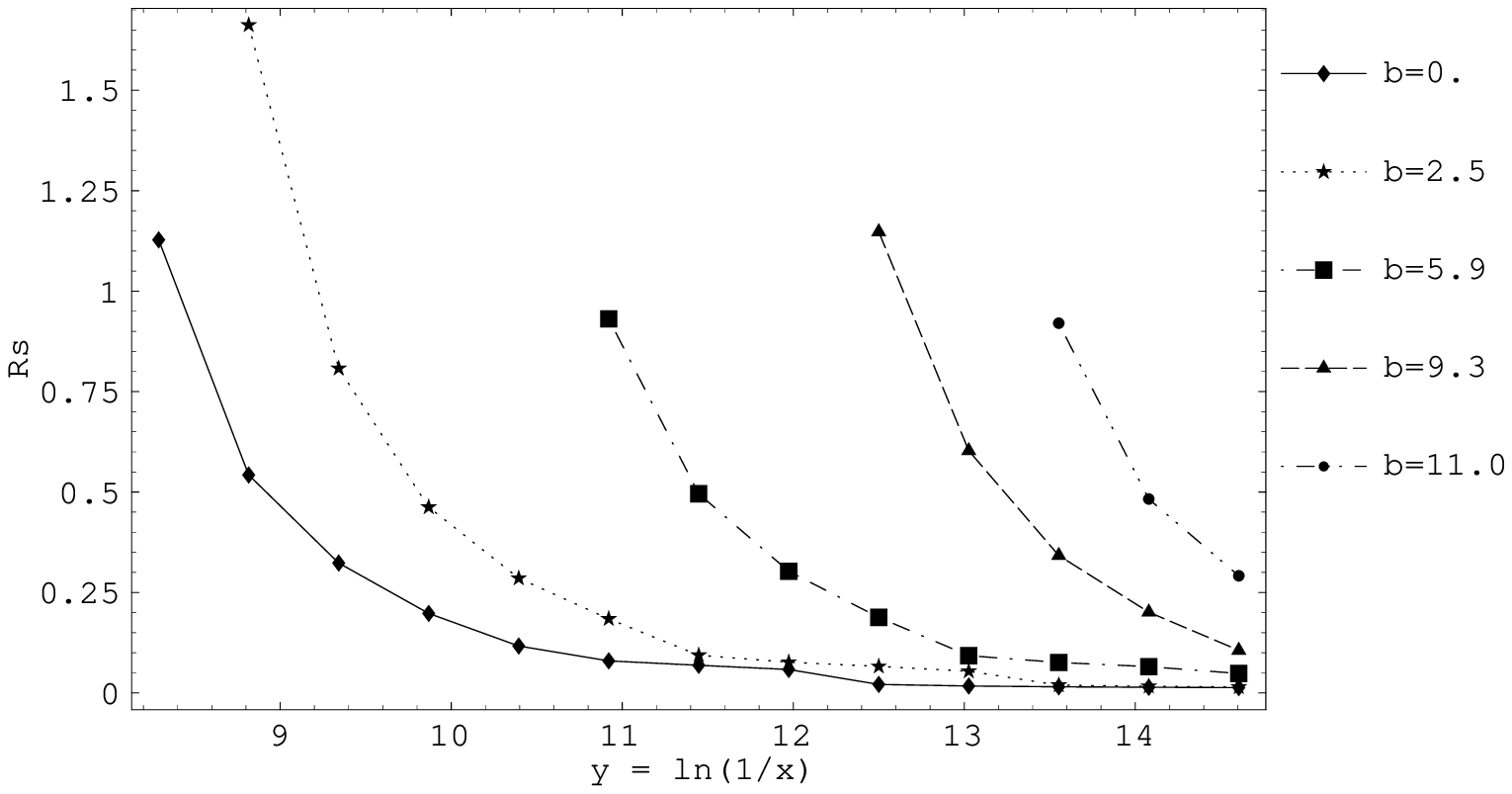, width=80mm} &
\epsfig{file= 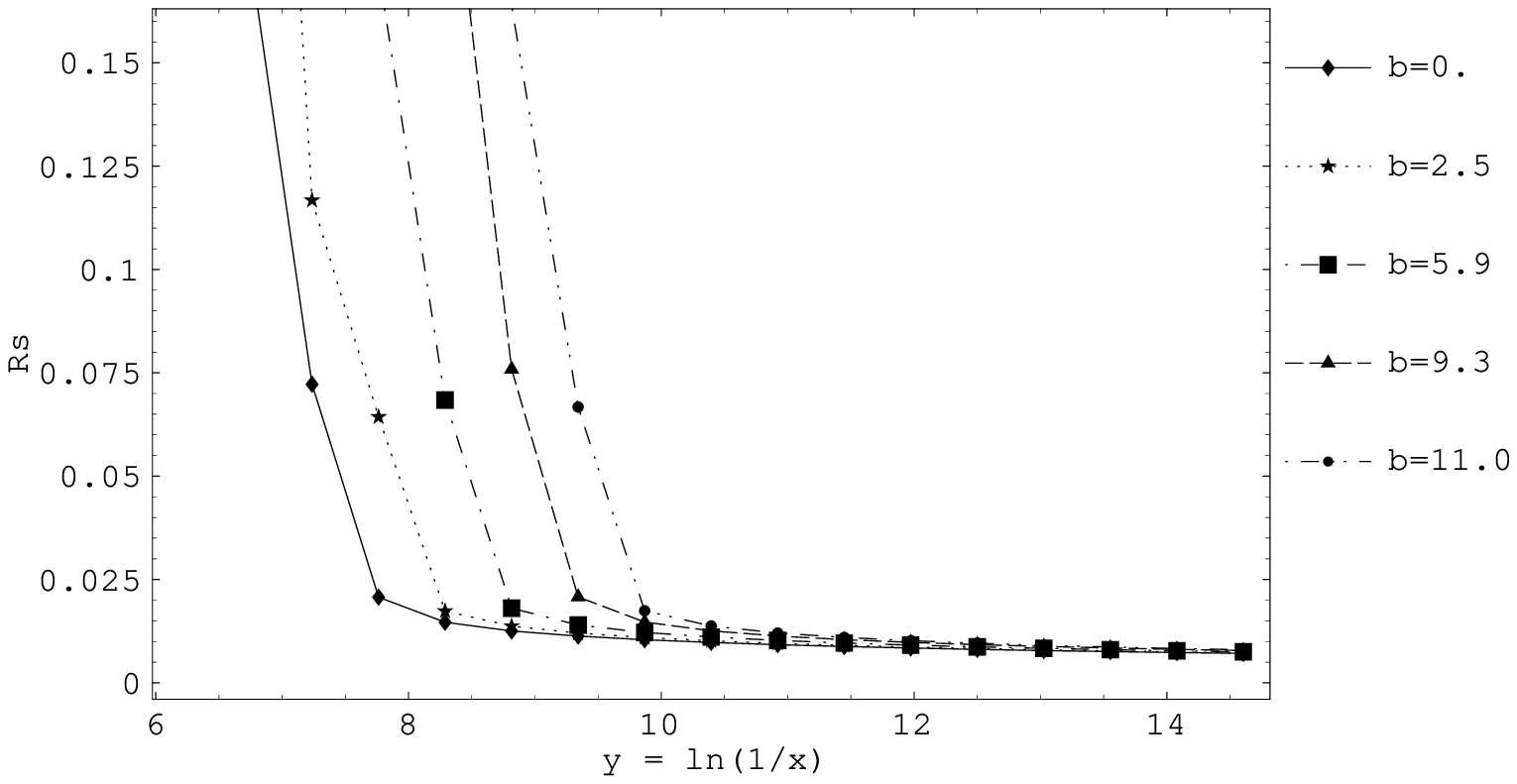, width=80mm}\\
\end{tabular}
\caption{The saturation scale as a solution to \eq{SATSCALE} of the amplitude
  for fixed and running QCD coupling, for different values of $y$ and 
$b$.}
\label{fig:satsc}
\end{figure}

 We shall now use our solution to the non-linear evolution equation
 to investigate
 saturation effects and scaling properties of the dipole-nucleon scattering
 amplitude. First, we calculated the saturation scale $r_s(y,b)$ as 
being the
 value of the dipole size, at which the amplitude 
\beq \label{SATSCALE}
N(r_s,y,b) = 1 \,-\,e^{-1} = 0.632
\eeq 
 \fig{fig:satsc} shows the solution to \eq{SATSCALE}. We recall, that for 
the
 case of fixed QCD coupling in the semiclassical solution (see Ref. \cite{BKL})
 we expect that at large $y$ and $b$ 
\beq \label{SATSCALE1}
Q^2_s(y,b)\,\,\propto\,\,e^{\lambda y}\,e^{-\frac{b}{R}},
\eeq
 where the estimate for the value of $\lambda $, is $\lambda = 4 
\alpha_S$ and
 $R$ is the radius of the target.

For the running QCD coupling constant,
the situation is more interesting (see Ref.
 \cite{GLR,MT,BKL}): (i) as function of $y$
\beq \label{SATSCALE2}
Q^2_s\,\,\propto\,\,e^{\sqrt{\lambda_R \,\,y}}
\eeq
in the region of $b \,\, <\,\,R\,\lambda_R \,y$ ;
and (ii) for $b \,\, >\,\,R\,\lambda_R \,y$ 
\beq \label{SATSCALE3}
Q_s\,\,\propto\,\,e^{ -\frac{b}{R}}\,\,;
\eeq
with $\lambda_R\,\,\approx \,\,10$.

To aid in understanding  the main features of the energy and impact 
parameter
 dependence of the saturation scale, we fitted the result for the saturation
 momentum using the following parameterizations, which reflect the above
 semiclassical  properties:
\bea 
\mbox{At fixed } b \mbox{:} & 
Q_s(y,b)\,\,=\,\,C_{y,1}\,e^{\lambda_F\,y}\,\,+C_{y,2}\,e^{\sqrt{\lambda_R\,y + a}} 
\label{QSFB} \\
\mbox{At fixed } y \mbox{:} &
Q^2_s(y,b)/4\,\,=\,\,\frac{C_{b,1}}{C_{b,2} +  \,e^{(\frac{(b + 
C_{b,4})}{C_{b,3}\,R})^2}} 
\label{QSFY}
\eea
%

\begin{figure}[t]
\begin{tabular}{c c}
Fixed $\as$ & Running  $\as$ \\
 ~ & ~\\
 \epsfig{file=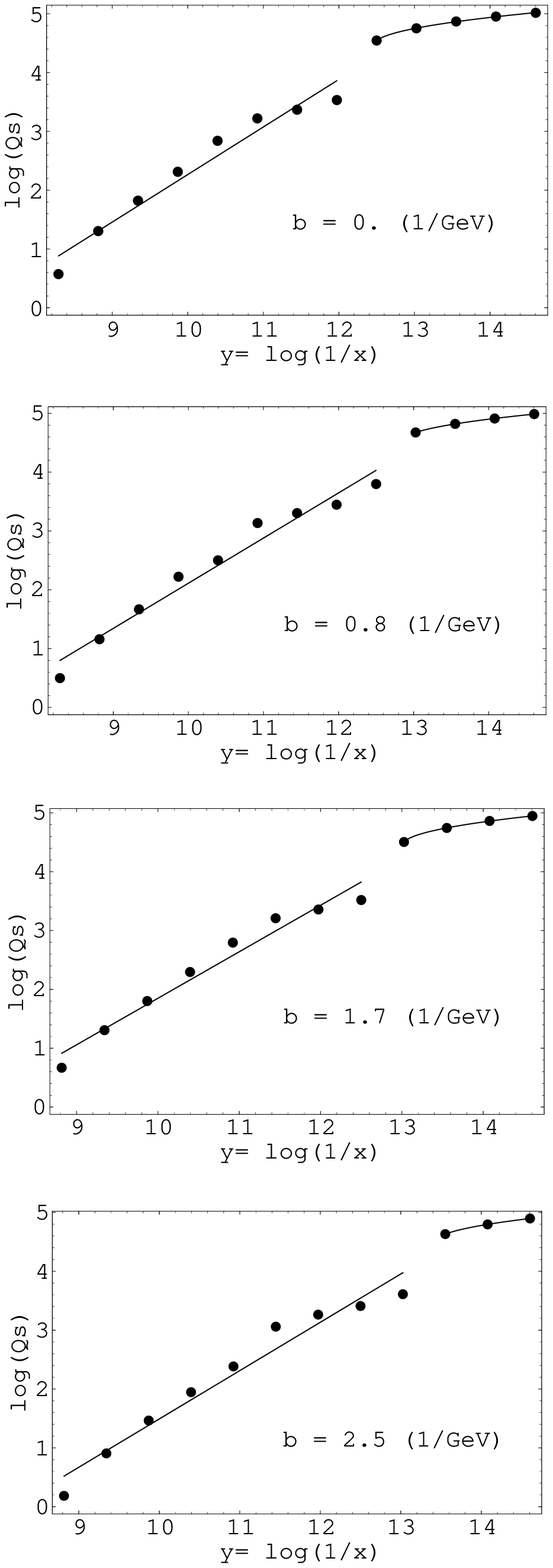,width=70mm} &
\epsfig{file=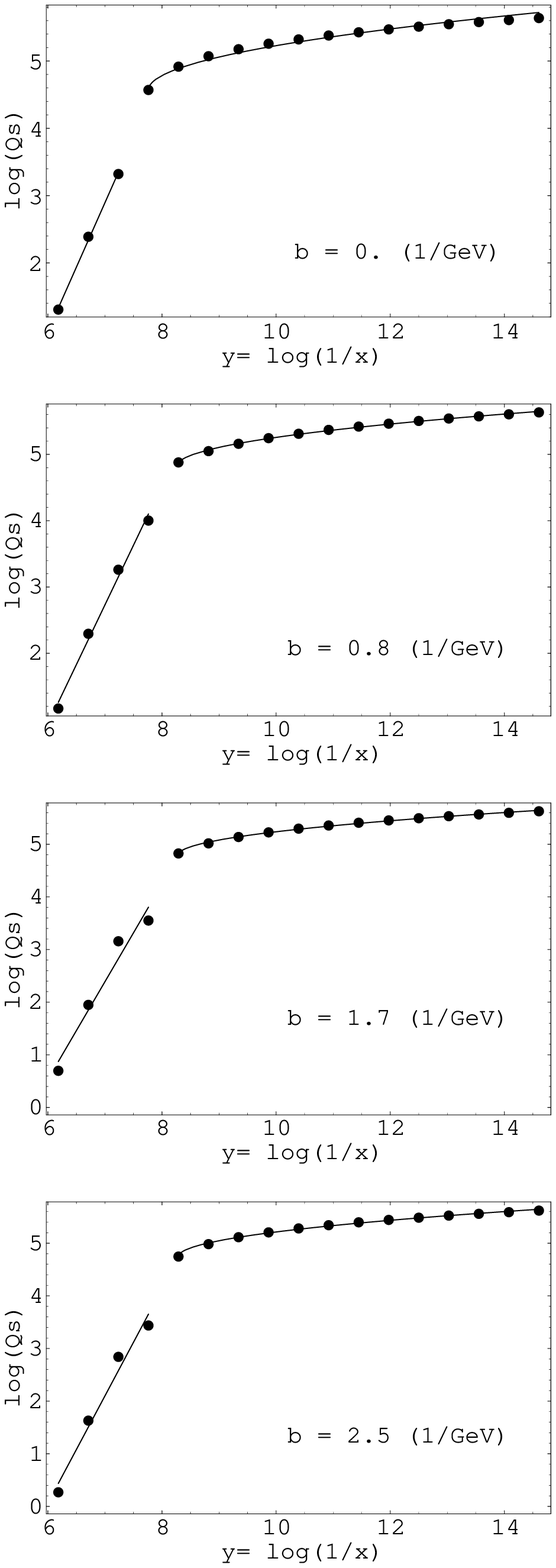,width=70mm}\\
\end{tabular}
\caption{The energy behavior of the saturation momentum $Q_s=2/r_s$ for 
fixed
and running QCD coupling. For small $y$ the curve is $\ln Q_s= a +
\frac{\lambda_F}{2}\,\,y$, while for large values of $y$ the curve
corresponds to $\ln Q_s= a +\frac{1}{2}\, \sqrt{\lambda_R + b}$. The 
results of  our 
calculations are presented as  circles whose magnitude reflect the size
of
possible numerical errors.   }
\label{fig:satscy}
 \end{figure}

In \fig{fig:satscy}
 we show how these parameterizations describe the calculated saturation 
momentum
 for fixed $b$. For running
 QCD coupling we used 
 the value of $\lambda_R$ = 0.26 to  0.30, which is
 much  smaller than the semiclassical prediction: 
$\lambda_R =
 10$. For fixed QCD coupling, we see an exponential growth with the value 
of
 $\lambda_F \approx$ 1.4 to 1.6, which is  approximately two times 
larger than
the theoretical 
prediction, $\lambda_F = 4 \as \approx 0.8$.

 Both $\lambda_F$ and $\lambda_R$ turn out to be independent  
 of $b$.

\begin{figure}[t]
\begin{tabular}{c c}
Fixed $\as$ & Running $\as$ \\
~ & ~\\
 \epsfig{file=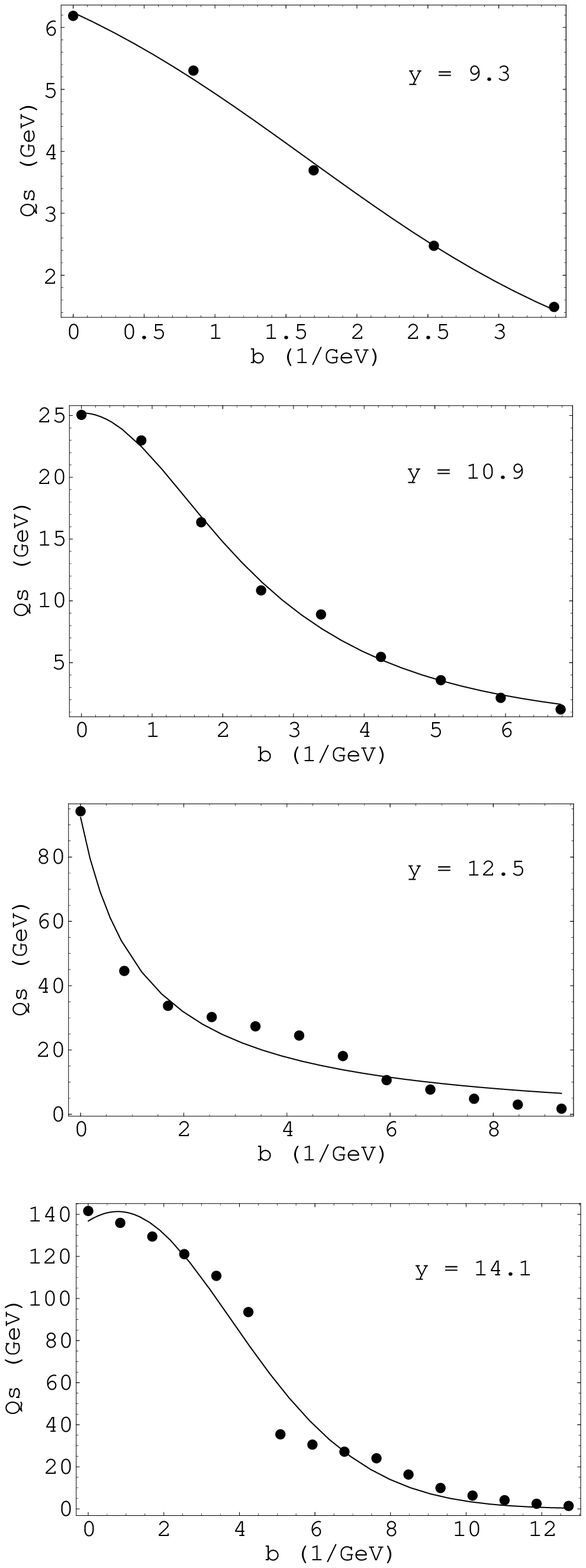, width=75mm} &
\epsfig{file=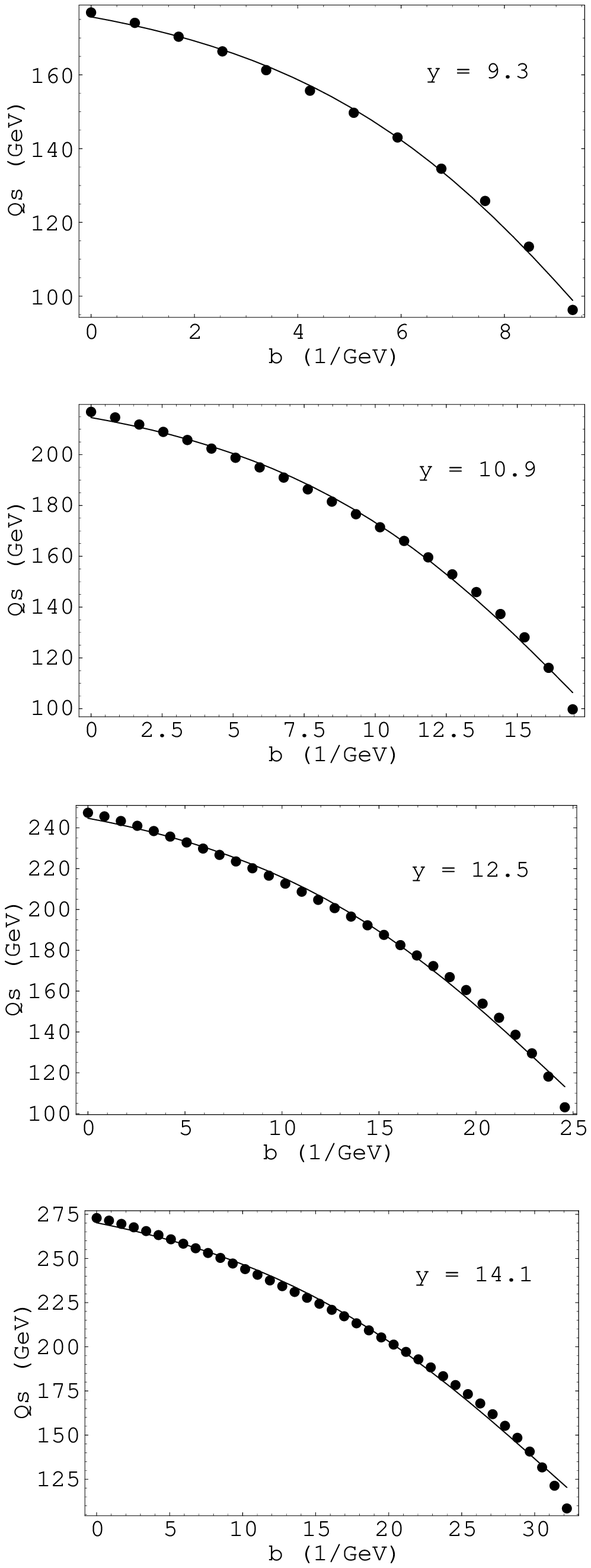, width=75mm}\\
\end{tabular}
\caption{The impact parameter ($b$) behaviour of the saturation momentum 
$Q_s=2/r_s$ for 
fixed
and running QCD coupling. The curve correspond to \eq{QSFB}.The results of 
our
calculations are presented as  circles whose magnitude reflect the size 
of
possible numerical errors.}
\label{fig:satscb}
\end{figure}


\begin{figure}[t]
\begin{tabular}{c c}
Fixed $\as$ & Running $\as$ \\
~ & ~\\
 \epsfig{file=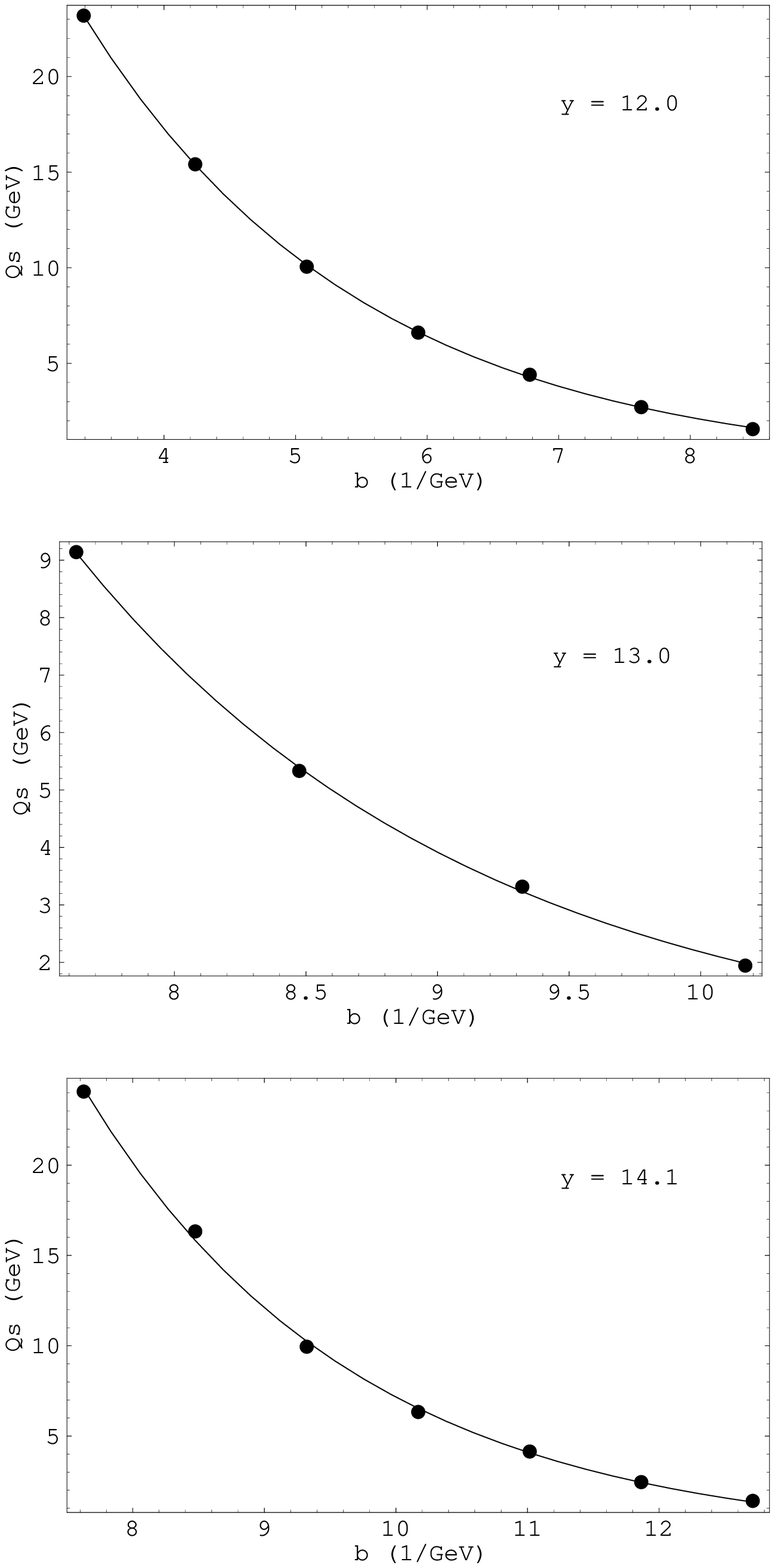, width=75mm} &
\epsfig{file=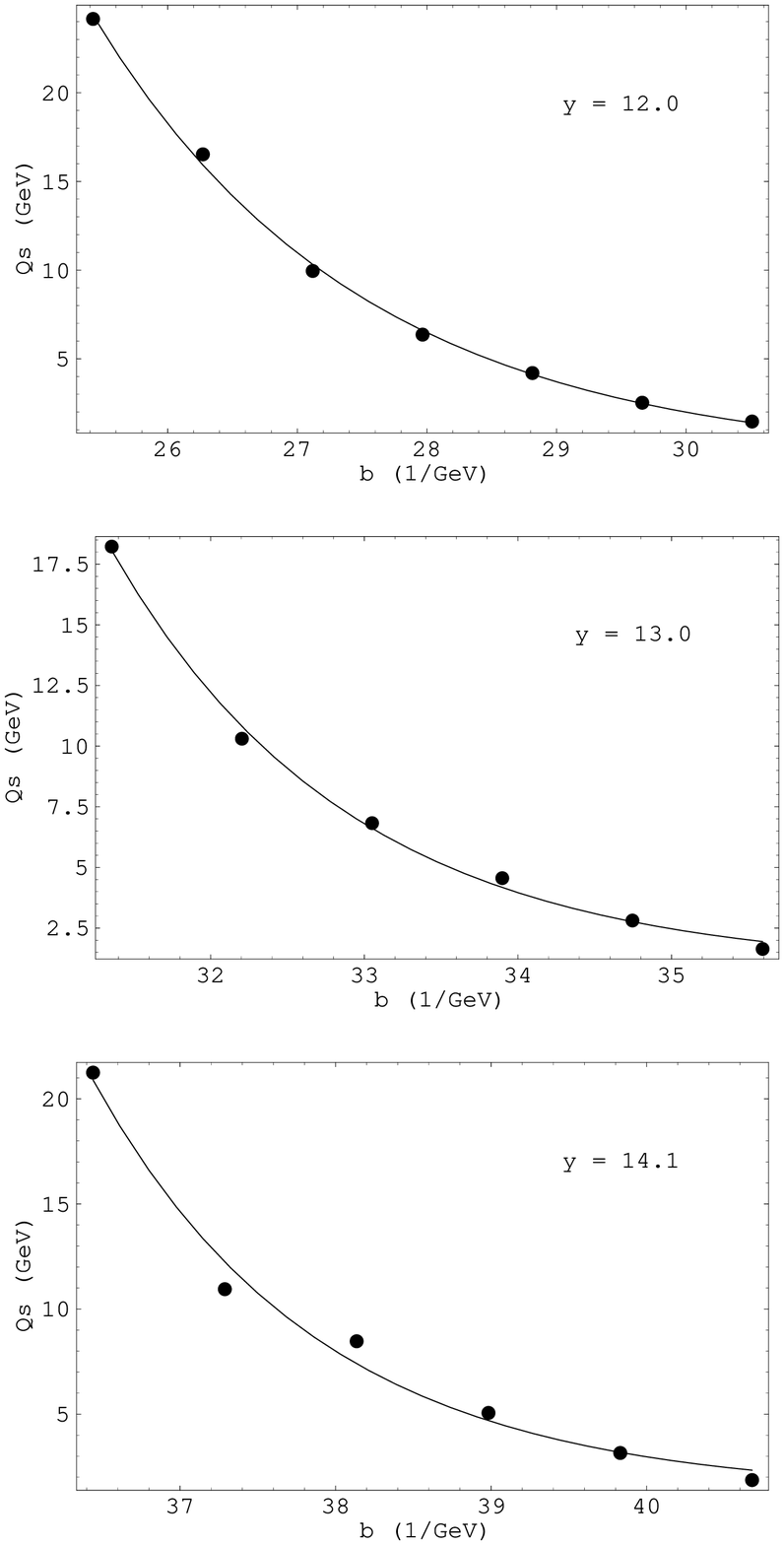, width=75mm}\\
\end{tabular}
\caption{Large  impact parameter ($b$) behaviour of the saturation 
momentum $Q_s$ for 
fixed
and running QCD coupling. The curve corresponds to $Q_s = C\,\,e^{ - 
m\,b}$
. The results of our
calculations are presented as  circles whose magnitude  reflect the size 
of
possible numerical errors.}
\label{fig:satscbtail}
\end{figure}

 \fig{fig:satscb}  shows the impact parameter
 behavior of the saturation scale. Both $y$ and $b$ behaviour of the 
saturation scale are in a qualitative agreement with the expectations mentioned 
above.
  The formula given in \eq{QSFY}
 describes  the main features of the $b$-behavior rather well. For fixed 
QCD coupling,
 we see that for a wide range of $y$ we have a Gaussian
 behavior, while for the case of running QCD coupling the decrease 
sets in, only in 
the
 region of very large $b$. However, the surprising fact is that even at fixed
 $\alpha_S$ for large $y$, the $b$-behavior tends to the same 
distribution 
as
 in the running $\alpha_S$ case. This observation requires   a 
qualitative explanation.  Note that at large $b$ in 
both 
cases, the impact parameter
 dependence appears to be Gaussian,  with  a 
radius of 
interaction  which increases with energy.
 In our parametrization the radius is equal 
to $R_{interaction} = C_3\,R$ and $C_3 \,\,\propto\,\,y$.

In \fig{fig:satscbtail} the large $b$ behaviour of the saturation scale
 is plotted.  
 $Q_s$ falls off as $e^{ - 0.5\,b}$ . One  expects
 such an exponential behaviour   from 
general properties of analyticity and crossing symmetry \cite{FROI}, and 
this behaviour supports 
our point of view that the initial conditions determine the large
 $b$ behaviour of the 
scattering amplitude.

\begin{figure}[t] 
\begin{tabular}{c c} 
Fixed $\as$  & Running $\as$ \\
 ~ & ~ \\
 \epsfig{file=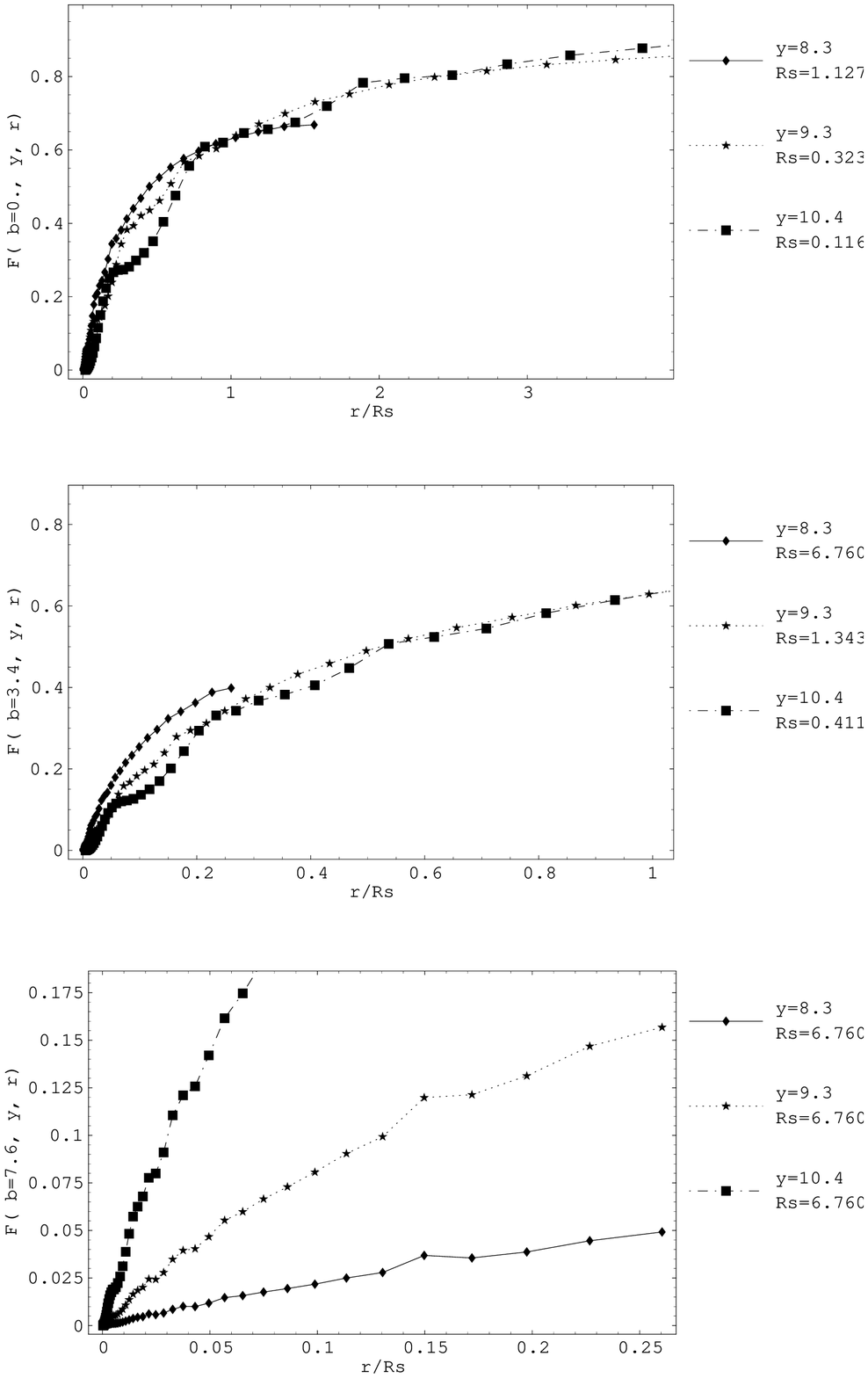,   width=65mm} &
 \epsfig{file=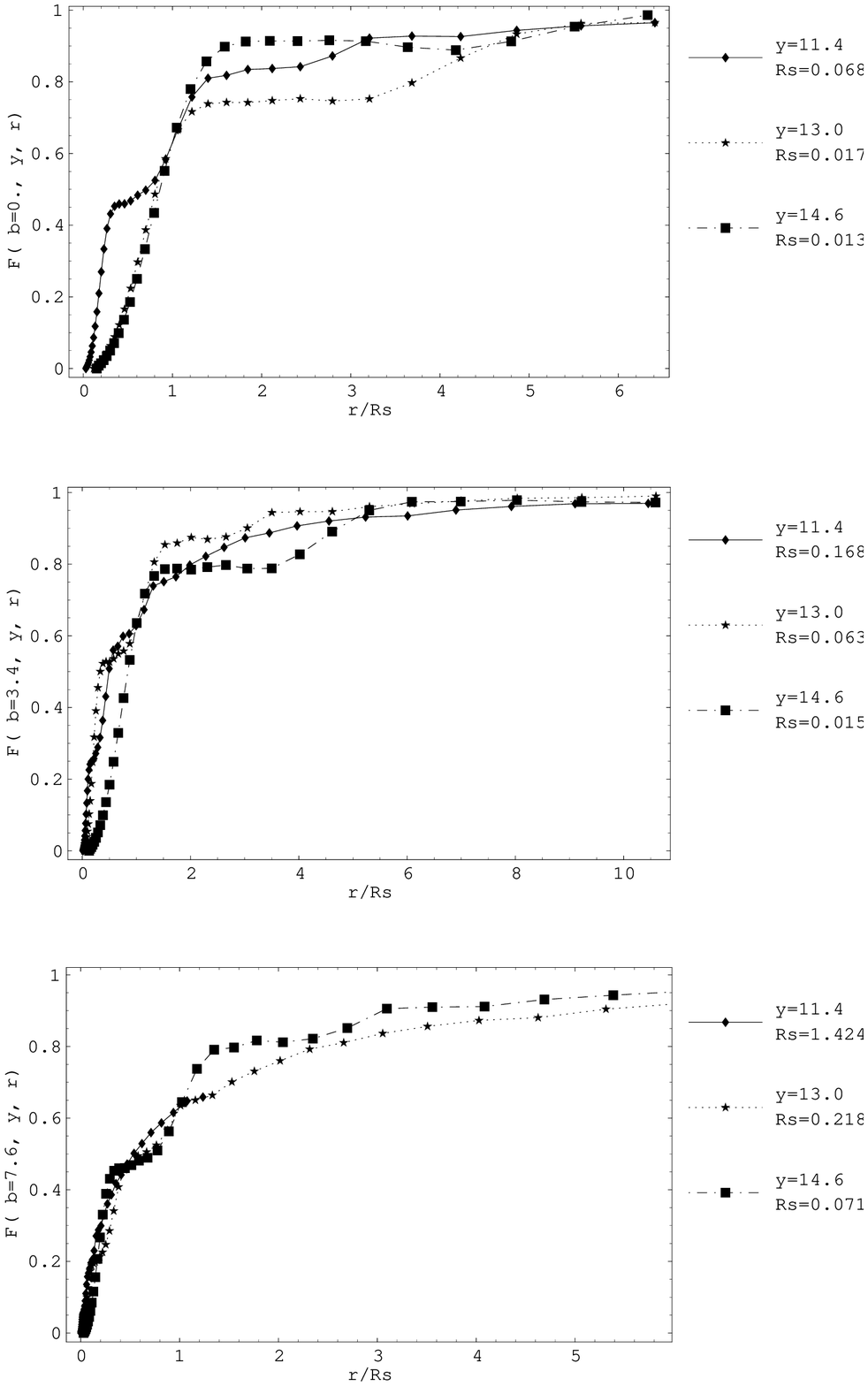,   width=65mm}\\
 \epsfig{file=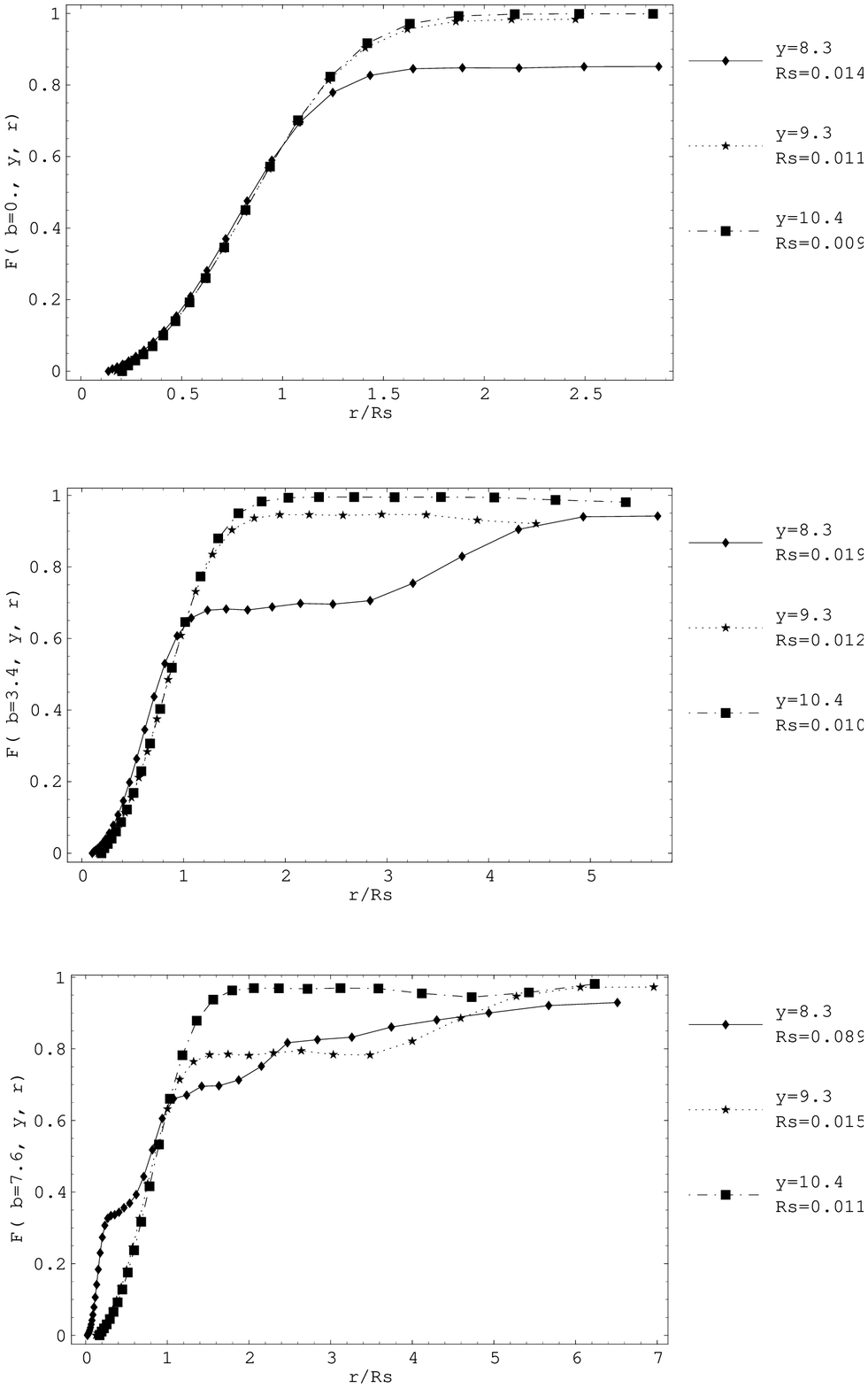, width=65mm} &
 \epsfig{file=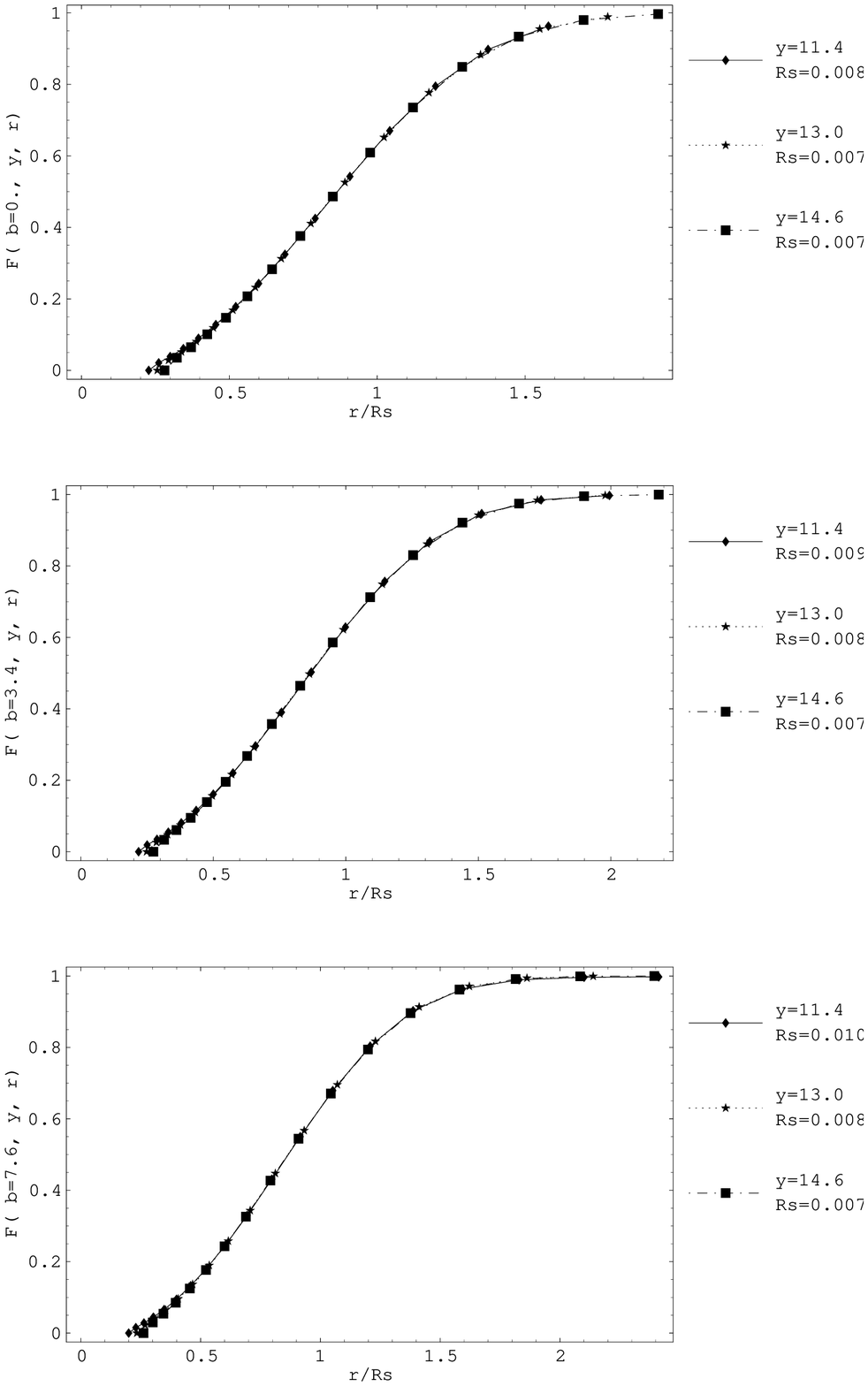, width=65mm}
 \end{tabular}
 \caption{The scaling behavior of the amplitude for fixed and running QCD
 coupling.}
 \label{fig:nscale}
 \end{figure} 

\subsection{Geometrical scaling}

In \cite{GLR,MV,GSC} it was shown that the dipole-nucleon scattering
 amplitude exhibits  scaling, which occurs due to the relationship
 between the $y$ and $r$ dependences of $N$.  This scaling behavior was
 confirmed in the approximate solution of the non-linear equation 
\cite{ML},
 without theoretical knowledge of the impact parameter behavior.  
More
 specifically, the scaling property implies that, in the kinematical region
 where saturation occurs, the amplitude is a function of one variable
 $\tau\equiv r Q_s(y,b)$, with $Q_s=2r_s^{-1}$.  

In \fig{fig:nscale} our solution is plotted as function of $\tau$.  
 A glance at this figure shows that the solution displays the scaling
 properties at large values of $y$, as was predicted by theory 
\cite{GSC}.  On
 the other hand, \fig{fig:nscale}  clearly displays that 
 geometrical scaling
 behavior is more pronounced,
for the running QCD coupling constant case.

  The improved geometrical scaling which we observe for the running
 coupling may, at first sight, be confusing,
since $\alpha_s$ contains a new dimensional parameter
 $\Lambda_{\mathrm{QCD}}$, which  in principle could spoil the scaling.
  However, in Ref.~\cite{BKL}
 it has been suggested that the running of the QCD coupling freezes at 
the
 saturation scale, thus recovering the broken geometrical scaling
 property. We interpret our numerical solution, as supporting this
  hypothesis.

Nevertheless, as the exact definition of the saturation region is a debatable
 point, we shall also
investigate the scaling property of $N$, using the ratio:
\begin{equation}\label{Rscaling}
{\cal R}=\frac{\dd N(y,r,b)/\dd y}{\dd\ln N(y,r,b)/\dd r}\,.
\end{equation}
 If scaling does exist, ${\cal R}$ should equal $\dd\ln Q_s(y,b)/\dd y$.
 Note, that the scaling property of $N$ implies that in the saturation 
region,
 ${\cal R}$ should not depend on $r$.  

\begin{figure}[t] 
\begin{tabular} {l l}
Fixed $\as$ & Running $\as$ \\
   &  \\
\epsfig{file=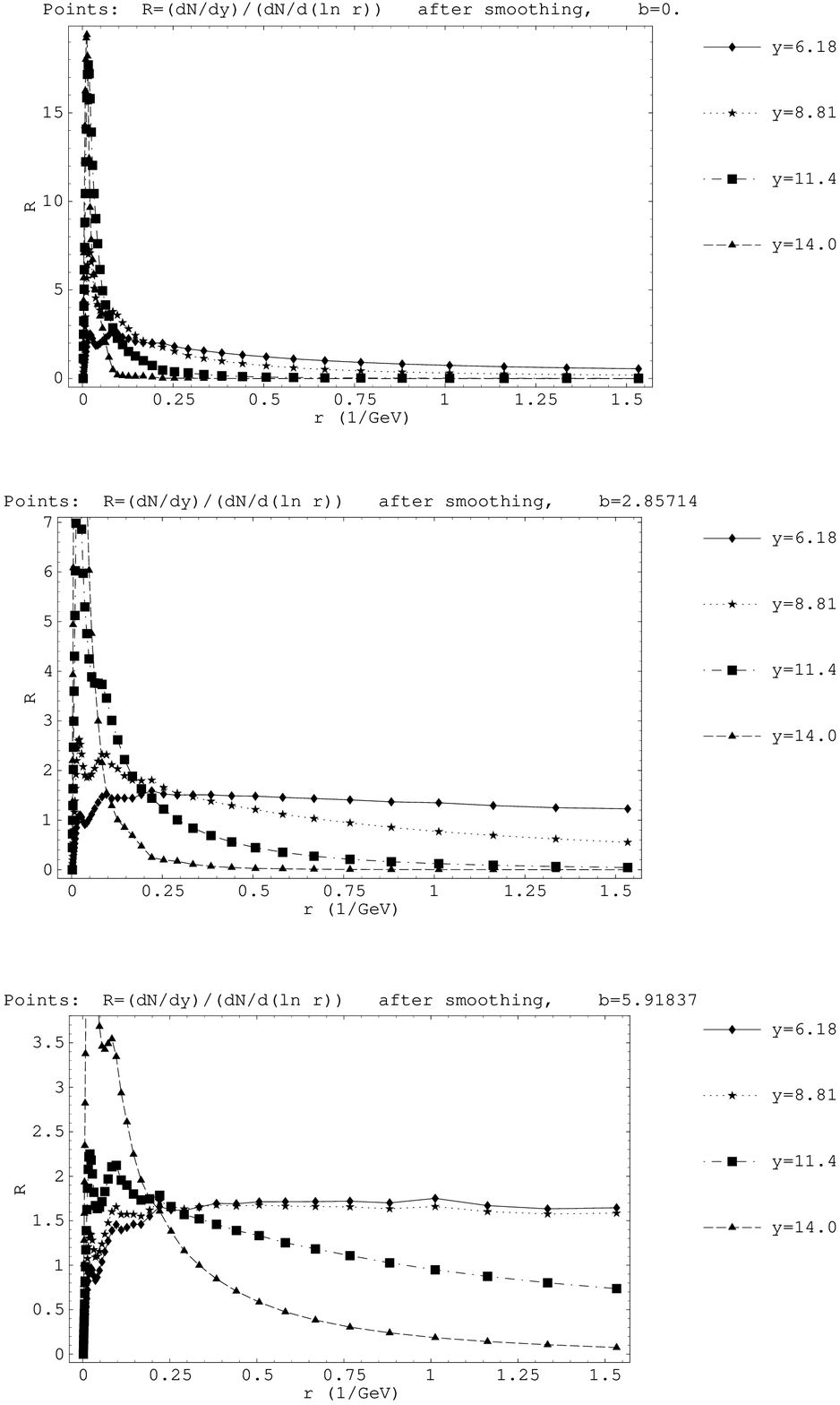,width=75mm} &
\epsfig{file=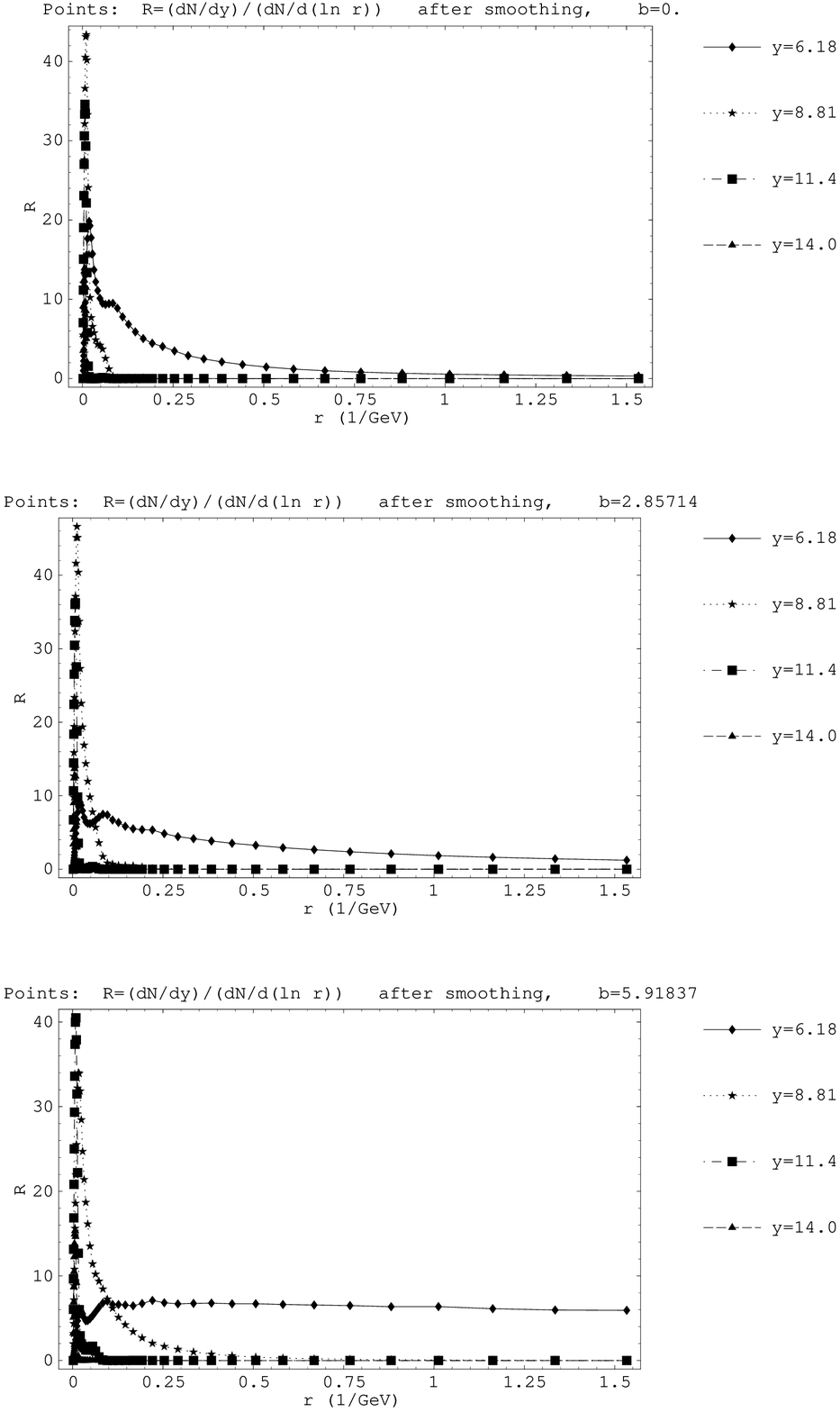,width=75mm}\\
\end{tabular}
  \caption{
Scaling ratio ${\cal R}$ as a function of dipole size $r$ for different 
values of $y$ and
$b$.}
\label{fig:scaling} 
\end{figure}

In \fig{fig:scaling} we present the ratio ${\cal R}$ as a function $r$, 
for
 different values of $y$ and $b$.  Although, as stated, our solution is
 only applicable at small values of $r$, we see that for long distances
 (for $r/R > 0.07$ with  $R=\sqrt{3.1}\,\gevm$)) ${\cal R}$
 is reasonably flat as a function
 of the dipole size, while for smaller values of $r$  the
ratio  falls steeply as a function of $r$.
 Once more,  the
geometrical scaling behaviour is more striking for the running  QCD
coupling case. The second interesting observation is that this
behaviour is concentrated at small $b$, while at large values of $b$ we
see a strong violation of geometrical scaling.

\subsection{Comparison with other solutions}

\begin{figure}[t] 
 \begin{center} 
\includegraphics[width=14cm]{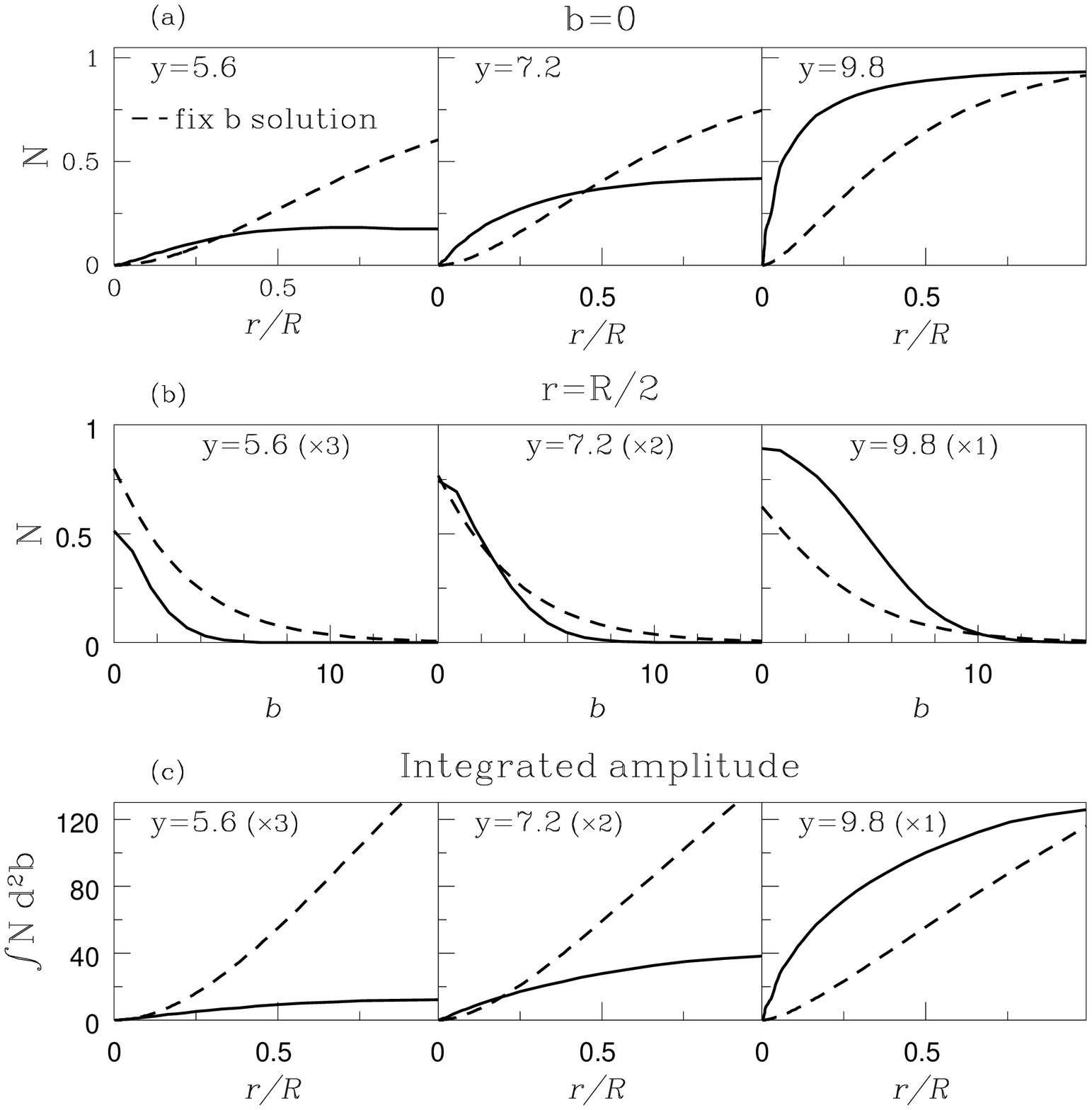}
\end{center} 
  \caption[]{\parbox[t]{0.80\textwidth}
{\small 
Comparison of  our solution (solid) and the fixed $b$ solution \cite{LL1}
 (dashed).  The comparison is for (a) the $r$-dependence of the amplitude at
 $b=0$; (b) the $b$-dependence of the amplitude at $r=R/2$; and (c) the
 integrated amplitude, $\int d^2b N(y,r,b)$ as a function $r/R$.}}
\label{fig:comp1} 
\end{figure}

The solution in \cite{LL1} was the  first step at solving the equation for fixed
 impact parameter.  This solution provided an excellent reproduction of the
 $F_2$ HERA data with $\chisquare=1$, using $R^2=3.1\,\gevs$.  In
 \fig{fig:comp1} we compare our solution (with the initial conditions of
 \fig{fig:initial}b) and the solution of \cite{LL1} (with the initial
 conditions of \fig{fig:initial}a).  \fig{fig:comp1}a shows the amplitude as
 a function of $r/R$.  The $r$-dependence of the solutions is substantially
 different.  We see that the $b$-dependence of the equation generates a rapid
 rise of the amplitude as a function of $r$.  

\fig{fig:comp1}b shows the $b$-dependence of the two solutions.  
 The impact parameter dependence of our present solution is the result of the
 evolution process (starting from the input of \fig{fig:initial}b), whereas
 the impact parameter dependence of \cite{LL1} was put by hand, according to
 the Glauber-Mueller formula $N(b)=1-\exp(-\kappa S(b))$, with
 $\kappa=-\ln(1-N(b=0))$.  The generated $b$-dependence differs
 from the ansatz of \cite{LL1}.  For low energy (\eg,
 $y=5.6$) the fixed $b$ solution extends to larger values of the impact
 parameter, whereas  at large energies ($y=9.8$) our present solution lies well above the
 fixed $b$ solution for values of $b$  about up to  $b=10$, where both solutions are small.  Only
 for intermediate energies and distances ($y\approx 7$, $r\approx R/2$), do the
 two approaches exhibit   similar behavior.

In \fig{fig:comp1}c we show how the two solutions differ in terms of the
 total cross section, \ie, the area under the curves of \fig{fig:comp1}b.
 Only for $y\approx 7$ and small distances do we find similarities between
 our present solution and the fixed $b$ solution of \cite{LL1}.

\begin{figure}[t]
 \begin{center}
\includegraphics[width=14cm, bb=0 160 580 710]{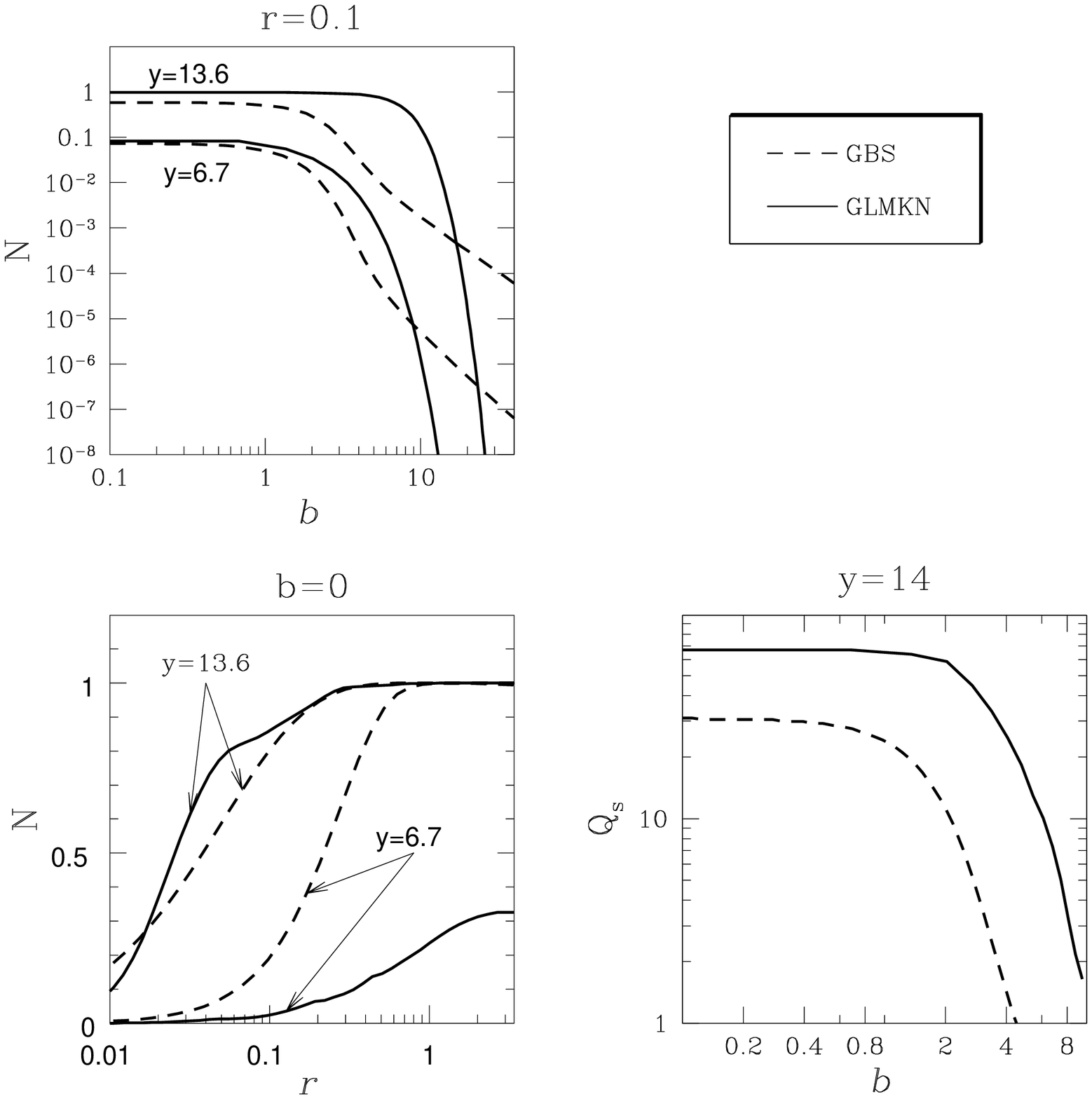}
\end{center}
  \caption[]{\parbox[t]{0.80\textwidth}
{\small
Comparison between  our solution (solid) and the solution of
\cite{GBS} (dashed).
}}
\label{fig:comp2}
\end{figure}

\fig{fig:comp2} shows a qualitative comparison of our solution and the
 solution of Ref.~\cite{GBS}.   Only a qualitative comparison can
 be made here, as (i) in \cite{GBS} there was an explicit $\cos\theta$
 dependence, where $\theta$ is the angle between $\rv$ and $\bv$, whereas our
 solution is an integrated quantity, $N(y,r;b)\equiv\int d\hrpar
 N(y,r;b;\hrpar)$; and (ii) there is a normalization difference, due to the
 different scales at which the dimensional variables are defined.

The origin of the last difference is the different initial conditions that
 were used.  In \cite{GBS} a pure Glauber-Mueller form for the initial
 conditions was assumed, using a profile $S(b)=10\exp{(-b^2/2)}$.  Our
 initial conditions, on the other hand, have a different shape (see
 \fig{fig:initial}b) and are based on parameters which were established for
 our fits to the HERA data \cite{LL1}.

 To enable us to make an illustrative comparison, we
 have used the following procedure:
\begin{enumerate}
\item To minimize the angular sensitivity,
 we  compare the two solutions only at low $r$ or low $b$.
\item We interpreted the input impact parameter dependence of \cite{GBS}
(\, i.e. $S(b)\propto\exp(-b^2/2)$\,), as $\exp{-b^2/R^2}$ with $R^2=2$.  
Recalling
that we use $R^2=3.1\gevs$ as the size of the target, we have rescaled 
$b$
according to $b^2\longrightarrow 2b/3$, hence matching the dimensional 
units
of \cite{GBS}.
 \item The shape of our initial input (see \fig{fig:initial}b) is
substantially different from the Gaussian of \cite{GBS}.  On the other
hand, we wish to compare the solution at $y>y_0$, where we assume that 
$N$
has a weak dependence (if any) on the shape of the initial conditions.  
Thus,
we have chosen to rescale $r$ so that after evolution of, say, two units 
of
rapidity\symbolfootnote[5]{Note that there is also a difference in the
definition of $y$: in \cite{GBS} $y_0=0$ and here
$y_0=-\ln{10^{-2}}\approx4.46$.  The values of $y$ in \fig{fig:comp2}
correspond to the latter definition}, the two solution will match at 
small
$b$ and fixed $r$. We found that a rescaling of $r\longrightarrow 2r$ is
sufficient to produce such matching.
\end{enumerate}

Referring again to \fig{fig:comp2} we see that as a function of $b$,
our solution decreases more rapidly.  In addition, we see that as a
function of $r$ and $b=0$, the large normalization difference between
the two solutions at $y-y_0=2$ diminishes at sufficiently large
$y-y_0=9$.  The $b$-dependence of the saturation scale $Q_s$ is
similar, but there is a considerably large normalization difference,
which may be due to disparities in the definition of the saturation
region, and the above mentioned dimensional parameters.

\section{Summary}
In this section we summarize the main features of our numerical
solution of the non-linear evolution equation.

\begin{itemize}

\item \quad It is shown that the non-linear equation in the region where
it's validity is guaranteed, leads to a solution which falls off as
\beq \label{FALLB}
N(r,b;y)\,\,\,\propto\,\,\,e^{- \frac{b}{R}}\,\,\,\,\,\,\mbox{for 
large}\,\,\,\,\,\, 
b
\,>\,\,\lambda\,y\,\,;
\eeq
\item \quad We showed that the inclusion of  dipoles of size
 larger than  the target ($R$), generate the power-like
tail in impact parameter distributions. This means that the power-like
behaviour is an artifact of using  the
non-linear equation, for calculating  the scattering amplitude in the
kinematic region, where this equation is not valid ( see Refs. \cite{IM}
for an approach that can describe this region);

\item \quad In our approach we modified the kernel of the non-linear 
equation by introducing $\Theta (R - r_1)\,\Theta (R - r_2)$.
We concur with the idea of Ref.\cite{KW} that to obtain the correct 
(exponentially decreasing) large impact parameter dependence, it is 
necessary to change the kernel of the non-linear equation.
 On the other hand, the non-linear 
equation was obtained implicitly assuming  that the size of the 
projectile dipole is 
much smaller than the size of the target. This was a misconception  
based on the semiclassical approach, and on the exact solution for the 
simplified double logarithmic kernel  for $r \,\ll\,R$, that  large 
dipoles will not appear in the intermediate stage of evolution 
\cite{LRREV,FIIM}. We demonstrated that the large $b$ tail which  behaves 
as $1/b^4$, stems entirely from the large dipole contribution, even for 
small 
$r$. The large $b$ tail gives a very small contribution to  the 
amplitude. One can compare \fig{fig:Nbfix} with the value of $\Delta N$ in 
\fig{fig:bdepdn}, to realize that $\Delta N/N$ is negligible for $b\, >\, 
25\,\, GeV^{-1} $.   Therefore, the 
statement of 
Refs. \cite{LRREV,FIIM} looks reasonable,  that the large $b$ tail which 
is  a 
problem 
in the 
pQCD approach, does not contributute to the main physical observables, 
including the total cross section.

\item \quad The value, energy and impact parameter dependence of the
saturation momentum are calculated. We find that at large values of
rapidity, the value of the saturation momentum   no longer depends on 
the impact
parameter $b$. A behaviour of this type
 was predicted for  $Q_s$ in the case of
running QCD coupling \cite{MT}, but was not expected for 
fixed $\alpha_S$;
\item \quad The geometrical scaling behaviour of the scattering amplitude
is discussed, and we show that the scattering amplitude 
can
be displayed as a function of one variable $\tau= r\,Q_s(y,b)$. 
 The geometrical scaling behaviour is more striking
for the running QCD coupling, than for fixed $\alpha_{s}$.
 An explanation for such behaviour
could be related to the idea \cite{BKL},  that the
running QCD coupling is frozen at the saturation scale. In this case the
smallness of $\alpha_S(Q_s)$ makes our calculations  more
reliable theoretically,
  and leads to  better geometrical scaling;
\item \quad The radius of interaction increases with energy, 
approximately
 behaving like
 $y$ $<b> \,\propto\,y$, in accordance with the Froissart
theorem \cite{FROI}. This result confirms that the large $b$
 behaviour stems from the non-perturbative scale 
introduced in the initial conditions;

 \item \quad The solution that we found is quite different from 
previous attempts, where an anzatz for $b$ dependence was assumed. We 
plan 
to
use our solution for developing a new procedure for a global fit to 
DIS data;

 \item \quad We  stress that the solutions for running and fixed 
$\as$ are quite different.
For running $\as$ we obtain a
 larger interaction radius (approximately twice as  large), steeper $y$ 
dependence, and  a larger value of the saturation scale;
\end{itemize}

\section*{Acknowledgments} 
We thank Anna Stasto for providing us with the numerical
calculations of Ref. \cite{GBS} and for useful discussions.

This research was supported in part by GIF grant $\#$ I-620-22.14/1999 and by
 Israeli Science Foundation, founded by the Israeli Academy of Science and
 Humanities.


\end{document}